%% file: main.tex
\documentclass[10pt,journal,compsoc]{IEEEtran}
\usepackage[utf8]{inputenc}
\ifCLASSOPTIONcompsoc
  \usepackage[nocompress]{cite}
\else
  \usepackage{cite}
\fi
\usepackage{adjustbox}
\usepackage{amsmath,amssymb,amsfonts}
\usepackage{algorithmic}
\usepackage{algorithm}
\usepackage{graphicx}
\usepackage{textcomp}
\usepackage{booktabs}
\usepackage{multirow}
\usepackage{float}
\usepackage{url}
\usepackage{enumitem}
\usepackage[hidelinks]{hyperref}
\usepackage{tabularx}
\usepackage[dvipsnames]{xcolor}
\usepackage{etoolbox}
\graphicspath{{figures/}}

\ifCLASSOPTIONcompsoc
  \usepackage[caption=false,font=normalsize,
              labelfont=sf,textfont=sf]{subfig}
\else
  \usepackage[caption=false,font=footnotesize]{subfig}
\fi

\let\oldthebibliography\thebibliography
\renewcommand{\thebibliography}[1]{%
  \oldthebibliography{#1}%
  \fontsize{7pt}{8pt}\selectfont
  \setlength{\itemsep}{0pt}%
}

\makeatletter
\long\def\@IEEEtitleabstractindextextbox#1{%
  \parbox{0.922\textwidth}{#1}}
\makeatother

\newcommand{\sys}{\textsc{TailRoll}}
\newcommand{\areal}{AReaL}

\providecommand{\cmark}{\ensuremath{\checkmark}}
\providecommand{\xmark}{\ensuremath{\times}}

\begin{document}

\title{\sys{}: Tail-Aware Rollout Acceleration for RL-Based LLM Post-Training}

\author{Niloofar~Gholipour,
        Marcos~Assuncao,
        Timothy~Yu,
        Gursimran~Singh,
        Rajkumar~Buyya,
        Julien~Gascon-Samson,
        Zhenan~Fan,
        and~Yong~Zhang
\IEEEcompsocitemizethanks{
\IEEEcompsocthanksitem N. Gholipour, M. Assuncao, and J. Gascon-Samson are with the \'Ecole de technologie sup\'erieure, Canada.\protect\\
E-mail: niloofar.gholipour.1@ens.etsmtl.ca, \{marcos.dias-de-assuncao, julien.gascon-samson\}@etsmtl.ca

\IEEEcompsocthanksitem T. Yu, Z. Fan, and Y. Zhang are with Huawei Technologies, Canada.\protect\\
E-mail: \{timothy.yu, zhenan.fan1, yong.zhang3\}@huawei.com

\IEEEcompsocthanksitem G. Singh is with RBC Borealis, Canada.\protect\\
E-mail: gursimran.singh@rbc.com

\IEEEcompsocthanksitem R. Buyya is with The University of Melbourne, Australia.\protect\\
E-mail: rbuyya@unimelb.edu.au
}
}

\markboth{IEEE Transactions on Parallel and Distributed Systems}%
{Gholipour \MakeLowercase{\textit{et al.}}: \sys{}: Tail-Aware Rollout Acceleration}

\IEEEtitleabstractindextext{%

\begin{abstract}
RL post-training of large language models (LLMs) is bottlenecked by rollout: heavy-tailed responses leave engines idle as each step waits for the longest response. Asynchronous training masks, but does not eliminate, idle time. We distinguish imbalance waste (uneven placement and decode contention) from straggler waste (waiting for the longest response). Tail length---not average load---often sets the step bound. We propose \sys{}, an online scheduler integrated into an open-source RL framework. It combines stateless in-flight migration and sufficiency-gated termination. Migration moves decoding requests between engines at chunk boundaries and resumes them with one re-prefill under a freeness-based, cost-aware tail-balancing policy. Termination stops a response when its GRPO group has enough completed samples for advantage estimation, masking its partial output from the training loss. Across ten configurations, \sys{} reduces rollout bubble ratio by 21--74\% and cross-engine KV-cache imbalance by 44--79\%. In end-to-end RL training, it cuts rollout time by 24--34\%, full step time by 21--22\% on long-CoT RL, and p95 request latency by 28--37\%, while raising generation throughput by up to 33\% with comparable or higher final reward.
\end{abstract}

}

\maketitle
\IEEEdisplaynontitleabstractindextext
\IEEEpeerreviewmaketitle

\input{sections/1.Introduction}

\input{sections/2.Background}

\input{sections/3.Motivation}
\input{sections/4.System-Overview}
\input{sections/5.TailRoll-Design}
\input{sections/6.Implementation}
\input{sections/7.Evaluation}
\input{sections/8.Related-work}

\input{sections/9.Discussion-Conclusion}

\bibliographystyle{IEEEtran}
\bibliography{ref}
\end{document}

%% file: sections/1.Introduction.tex
\IEEEraisesectionheading{\section{Introduction}
\label{sec:intro}}

\IEEEPARstart{R}{einforcement} learning (RL) post-training improves pretrained large language models (LLMs) using rewards assigned to model-generated responses. In GRPO~\cite{shao2024deepseekmath}, for example, the policy samples multiple responses per prompt, assigns rewards, and updates toward higher-reward responses. Each iteration comprises rollout generation, response scoring, policy updates, and weight distribution to inference engines. Rollout accounts for 50--76\% of training-step time across our workloads (Fig.~\ref{fig:findings} in Sec.~\ref{sec:motivation}), consistent with findings in other RL systems~\cite{zhong2025streamrl,gao2025rollpacker,he2025rhymerl}. Unlike training, which processes tokens in parallel batches, rollout generates tokens sequentially, making engine idle time a significant source of accelerator waste.

\begin{figure}[t]
  \centering
  \includegraphics[width=0.88\columnwidth]{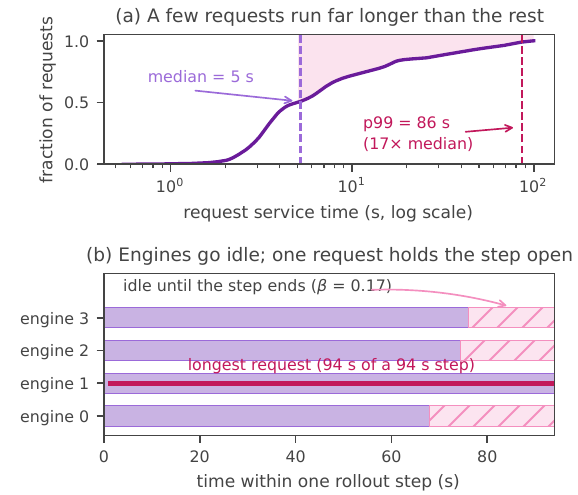}
  \caption{\scriptsize The problem (multimodal-heavy, Qwen3-14B, four engines). (a) Per-request service time over 50 steps: median 5\,s, p99 86\,s. (b) One step: engines finishing early remain idle until the longest request (94\,s) completes.}
  \label{fig:intro-problem}
\end{figure}

Response lengths in RL workloads are highly variable, with a small fraction of long-tail trajectories running orders of magnitude longer than the median~\cite{gao2025rollpacker,he2025rhymerl,qin2026seer}. Because training waits for the entire batch, engines that finish their assigned requests early remain idle while others decode stragglers. Fig.~\ref{fig:intro-problem} illustrates this behavior: a few requests dominate service time (a), leaving early-finishing engines idle (b). We quantify this loss using the \emph{bubble ratio} $\beta$, the fraction of engine time spent idle during rollout, which ranges from 0.17 to 0.57 across our workloads. Unlike cluster-level scheduling~\cite{chen2023deepboot,liu2024sampling}, recovering this capacity requires decisions within a rollout step.

Existing approaches address different aspects of rollout inefficiency. \emph{Asynchronous training} (\areal{}~\cite{fu2025areal}, AsyncFlow~\cite{han2025asyncflow}, StreamRL~\cite{zhong2025streamrl}) disaggregates rollout and training, allowing engines to process requests concurrently without waiting for the trainer. However, it does not eliminate idle time within a rollout collection window. \emph{Length-aware scheduling} (RollPacker~\cite{gao2025rollpacker}, SortedRL~\cite{zhang2026sortedrl}, RhymeRL~\cite{he2025rhymerl}) estimates response lengths and places requests before generation, but cannot adapt to imbalance emerging during decode. \emph{Chunk-based scheduling}, exemplified by Seer~\cite{qin2026seer}, reassigns generation chunks through a shared KV-cache pool and uses sibling-length estimates to improve synchronous RL throughput. Despite acting during decode, it leaves three gaps: request counts do not capture differences in remaining work; its synchronous setting does not address idle time within asynchronous collection windows; and its placement decisions cannot stop a response that determines the step's completion. Thus, existing approaches do not directly address whether a critical-path response still needs to be generated.

We establish a lower bound on rollout completion time under fixed request service times: the maximum of total work divided by the number of engines and the longest individual request's service time (Sec.~\ref{sec:formulation}). Both terms can be computed from controller-recorded spans. Across our workloads, observed step times are within 5\% of this bound, usually within 1\%, with the longest request always determining the bound. Moreover, an oracle dispatcher given exact output lengths before generation finishes within 0.2\% of the stock dispatcher. These results show that dispatch-time placement alone offers little improvement when request service times remain fixed.

This analysis separates idle time into two sources: \emph{imbalance waste} ($W_{\mathrm{imb}}$), caused by uneven request placement and decode contention, and \emph{straggler waste} ($W_{\mathrm{strag}}$), caused by waiting for the longest response. They require different remedies. Migration can reduce the service time of a straggler by moving other requests off its engine, reducing batch depth and improving the remaining requests' decode rates. However, only termination can eliminate the need to generate the remaining tokens of a response.

We propose \sys{}, an online rollout scheduler integrated into an open-source RL framework for synchronous and asynchronous training. To reclaim imbalance waste, \sys{} introduces \emph{in-flight request migration}: a background monitor tracks engine load and KV-cache pressure, while a cost-aware planner moves eligible requests between engines at chunk boundaries. The destination reconstructs request state from the prompt and generated tokens through a single prefill before resuming decode. Cost-benefit admission, destination reservations, hop limits, and post-migration verification maintain stable scheduling as load evolves. To address straggler waste, \sys{} introduces \emph{sufficiency-gated termination}. Since GRPO responses contribute to a shared group-relative advantage estimate, the scheduler stops a straggler once at least $k_{\min}$ of its $n$ siblings have completed, retaining its partial output with the corresponding training loss masked. Our 300-step experiments show comparable training reward (Sec.~\ref{sec:eval-term}).

\textbf{Contributions.} This paper makes five contributions:
\begin{enumerate}[leftmargin=*,nosep]
\item We identify and quantify two sources of rollout idle time: placement-reclaimable imbalance and straggler-induced critical-path waste. Per-step span accounting and an oracle dispatcher across six workloads establish the limits of dispatch-time placement.
\item We design and implement \sys{}, an in-flight scheduling framework integrated into \areal{} through an observe--decide--execute control plane, enabling mid-generation request migration without modifying inference engines or the trainer.
\item We develop a multi-stage, cost-aware migration policy combining freeness-based imbalance detection, tail-aware source--target pairing, migration cost-benefit filtering, and bounded execution with reservations, hop limits, and post-move verification.
\item We introduce sufficiency-gated termination, which stops critical-path stragglers once their GRPO groups have enough completed siblings, extending scheduling from deciding where requests execute to whether their remaining tokens need to be generated.
\item We evaluate \sys{} on Ascend 910C NPUs with Qwen3-14B and Qwen3-32B across long-CoT RL and production-derived workloads. It reduces rollout bubble ratio by 21--74\% and cross-engine KV-cache imbalance by 44--79\%, cuts rollout time by 24--34\% and end-to-end step time by 21--22\%, reduces p95 TTFT by 71--88\% and p95 request latency by 28--37\%, and increases generation throughput by up to 33\%, while preserving reward over 300-step runs.
\end{enumerate}

The remainder of this paper is organized as follows. Sec.~\ref{sec:background} reviews RL post-training and inference engines; Sec.~\ref{sec:motivation} presents our measurement study; Sec.~\ref{sec:overview} describes the architecture; Sec.~\ref{sec:design} details the idle-time model, migration, scheduling policy, and termination; Sec.~\ref{sec:impl} describes implementation and methodology; Sec.~\ref{sec:eval} presents evaluation results; Sec.~\ref{sec:related} discusses related work; and Sec.~\ref{sec:discussion} concludes the paper.

%% file: sections/2.Background.tex
\section{Background}
\label{sec:background}

\subsection{RL Post-Training and the Rollout Barrier}

A GRPO-style RL step~\cite{shao2024deepseekmath} samples $n$ responses for each of $B$ prompts, scores them, normalizes rewards within each group to obtain advantages, and applies a clipped policy-gradient update. In synchronous training, the update waits for all $B{\cdot}n$ responses, making the longest response determine rollout completion~\cite{gholipour2026rollout}. Two properties motivate our termination mechanism. First, each response's advantage depends on rewards across its siblings. Second, generated trajectories serve training rather than user delivery. Thus, once enough siblings have completed to support the group estimate, continuing a straggler may add little value while delaying the step. Existing approaches relax the full-batch barrier differently: APRIL~\cite{zhou2025april}, Kimi~k1.5~\cite{team2025kimi}, and verl's partial rollout~\cite{sheng2025hybridflow} defer unfinished trajectories to later steps, while Selective Rollout~\cite{Zheng2026ActOnlyWhenItPays} and ESPO~\cite{Li2026ESPO} use static per-sample stopping or selection rules. None combines a system-level critical-path condition with live group-sufficiency gating while retaining partial trajectories through loss masking. Sec.~\ref{sec:termination} presents our approach.

\subsection{Inference Engines: Batching and Prefill}

LLM inference engines such as vLLM and SGLang use continuous batching to serve concurrent requests. Prior work studies GPU sharing, elastic batch scheduling, and asynchronous operator parallelism~\cite{han2024inss,qi2024elasticbatch,lin2025top}, primarily targeting serving-time latency rather than RL step completion. At each decode iteration, an engine generates one token per active request, sequentially across each request's output. Consequently, per-request decode rate depends on both batch depth and context length. In our measurements, the decode rate at batch depth 256 is half that at depth 1 (Sec.~\ref{sec:eval-bound}). In contrast, known token sequences can be processed in parallel during \emph{prefill}. Since prefill is cheaper per token than decode in our measurements (Sec.~\ref{sec:design}), a destination engine can reconstruct a migrated request's state from its prompt and generated tokens rather than receive its KV cache.

\subsection{Asynchronous RL Systems and the \areal{} Prototype}

Asynchronous RL systems~\cite{fu2025areal,han2025asyncflow,zhong2025streamrl} separate rollout and training across device pools, allowing the trainer to consume responses generated by bounded-staleness policy versions. This hides rollout latency from the training critical path but does not eliminate imbalance within the rollout pool. In our \areal{}~\cite{fu2025areal} prototype, a round-robin router dispatches requests to inference engines, where they remain pinned for their lifetime. Consequently, overloaded engines accumulate queues while others become under-utilized, even with asynchronous execution. Sec.~\ref{sec:motivation} quantifies this effect across the tested staleness levels (Insight~4, Fig.~\ref{fig:findings}).

\subsection{Moving a Request After Decode Has Started}
\label{sec:bg-migration}

GPU job migration has been studied at the cluster, application, and container levels~\cite{prades2019rcuda,wang2023dcuda,kennedy2023avec}, but these approaches do not relocate individual decoding requests. At the serving level, Llumnix~\cite{sun2024llumnix} transfers KV blocks between vLLM instances during decode. Seer~\cite{qin2026seer} instead uses fixed-size generation chunks backed by a shared KV-cache pool, allowing requests to move between instances without recomputing their prefixes. \sys{} adopts a stateless alternative: it transfers only the prompt and generated tokens, allowing the destination to rebuild request state through one parallel prefill. This avoids KV-cache transfer, shared storage, and dependence on engine internals, using the controller's existing HTTP path instead. Its trade-off is prefix recomputation, so migration is admitted only when the remaining decode work justifies the prefill cost (Secs.~\ref{sec:mechanism} and~\ref{sec:eval-overhead}).

%% file: sections/3.Motivation.tex
\section{Motivation: Long-Tail Responses Drive Rollout Waste}
\label{sec:motivation}

This section gives the insights that lead to the design in Sec.~\ref{sec:design}. The measurements below use the same runs as the evaluation in Sec.~\ref{sec:eval}: \areal{} on Ascend 910C NPUs, Qwen3-14B and Qwen3-32B, four rollout engines, and the workloads in Table~\ref{tab:workloads} (full setup in Sec.~\ref{sec:eval-setup}). The two multimodal rows replay response-length distributions from the Azure multimodal inference trace~\cite{azuretrace} (58\% of source requests carry at least one image); multimodal-normal uses the trace as-is, while multimodal-heavy is a longer-tailed pool derived from the same source.

\begin{table}[t]
    \centering
    \caption{\scriptsize Workloads (baseline runs, synchronous, $N{=}0$). Lengths are mean per-step p50/p99 response lengths; $\beta$ is Eq.~\eqref{eq:bubble}.}
    \label{tab:workloads}
    \setlength{\tabcolsep}{5pt}
    \small
    \resizebox{0.92\columnwidth}{!}{%
    \begin{tabular}{@{}lcc|c|cc@{}}
    \toprule
    Workload & p50 & p99 & \multicolumn{1}{c|}{p99/p50}
    & $\beta$ & \shortstack{Rollout\\s/step} \\
    \midrule
    Multimodal-normal & 101   & 1\,116 & 11.3 & 0.312 & 58.3   \\
    Multimodal-heavy  & 77    & 2\,664 & 36.9 & 0.171 & 90.0   \\
    Straggler pool    & 40    & 395    & 9.8  & 0.331 & 29.6   \\
    CodeAgent         & 68    & 1\,607 & 23.8 & 0.305 & 77.1   \\
    Multitask         & 366   & 6\,269 & 17.9 & 0.249 & 237.6  \\
    AMC12 (8K cap)    & 5\,434 & 8\,192 & 1.6 & 0.175 & 644.7  \\
    \bottomrule
    \end{tabular}}
    \end{table}

\textbf{Insight 1: rollout is the bottleneck stage of LLM RL post-training.}
On multimodal-heavy, rollout takes 50\% of the measured step time for Qwen3-14B and 55\% for Qwen3-32B. On the workloads with longer responses it takes more: 66\% on AMC12 and 76\% on multitask. The share grows with response length, because generation is sequential decoding while the training stages process the same tokens in parallel batches. Rollout is therefore not a small cost that a faster trainer can hide.

\begin{figure*}[t]
\centering
\begin{minipage}[t]{0.235\textwidth}
\centering
\includegraphics[width=\linewidth]{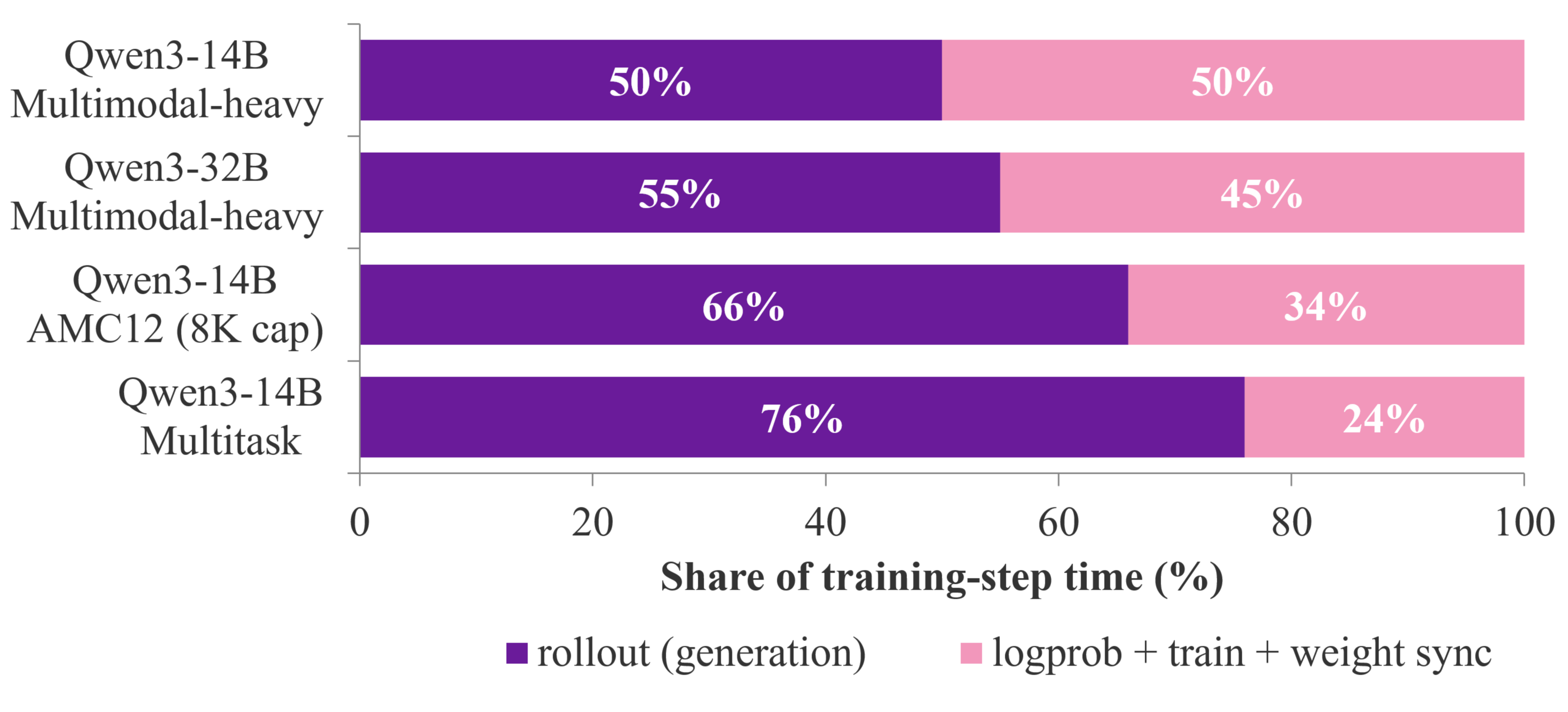}
\par\scriptsize (a) Rollout share of the step
\end{minipage}\hfill
\begin{minipage}[t]{0.235\textwidth}
\centering
\includegraphics[width=\linewidth]{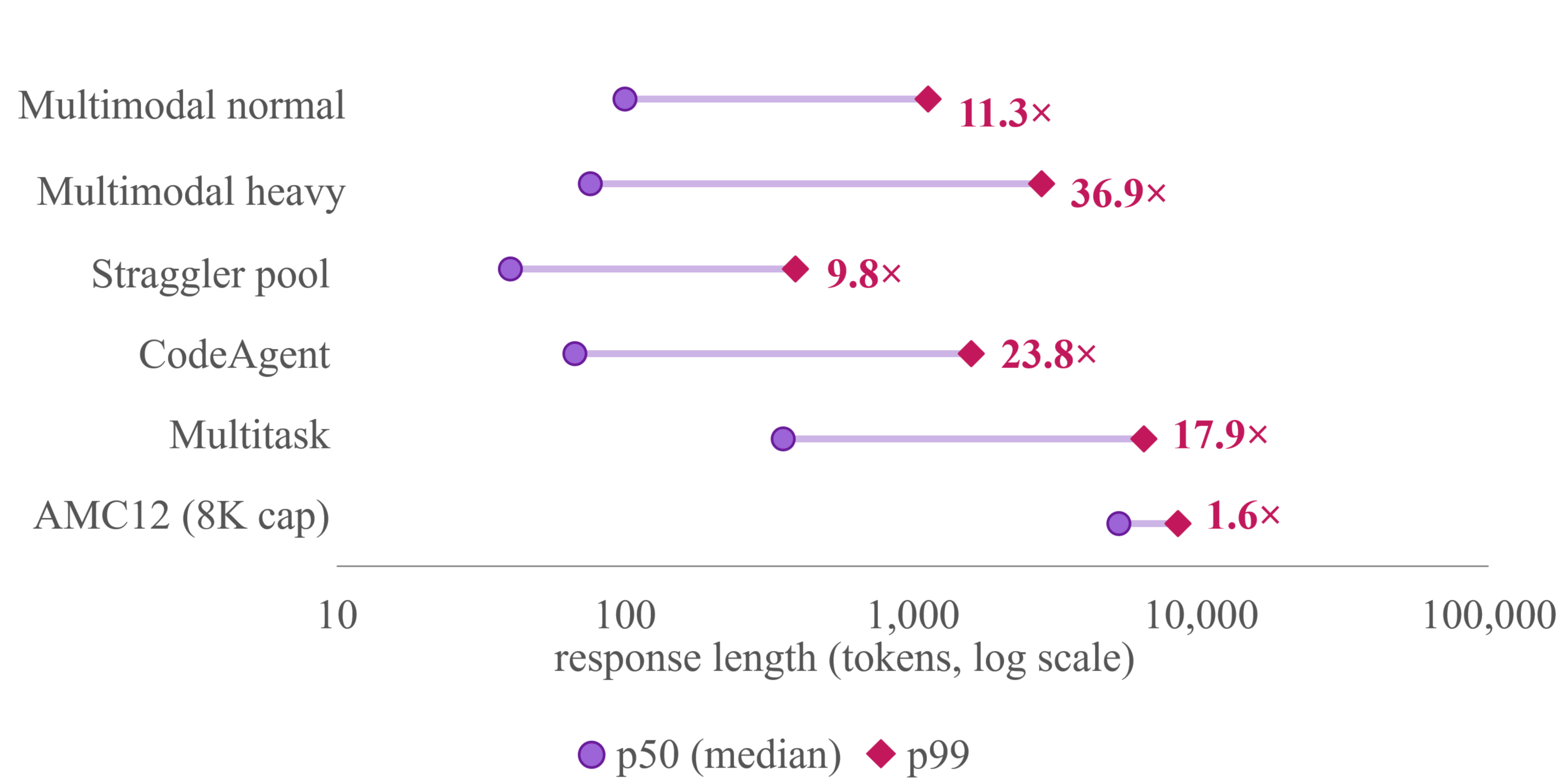}
\par\scriptsize (b) p50/p99 response lengths
\end{minipage}\hfill
\begin{minipage}[t]{0.235\textwidth}
\centering
\includegraphics[width=\linewidth]{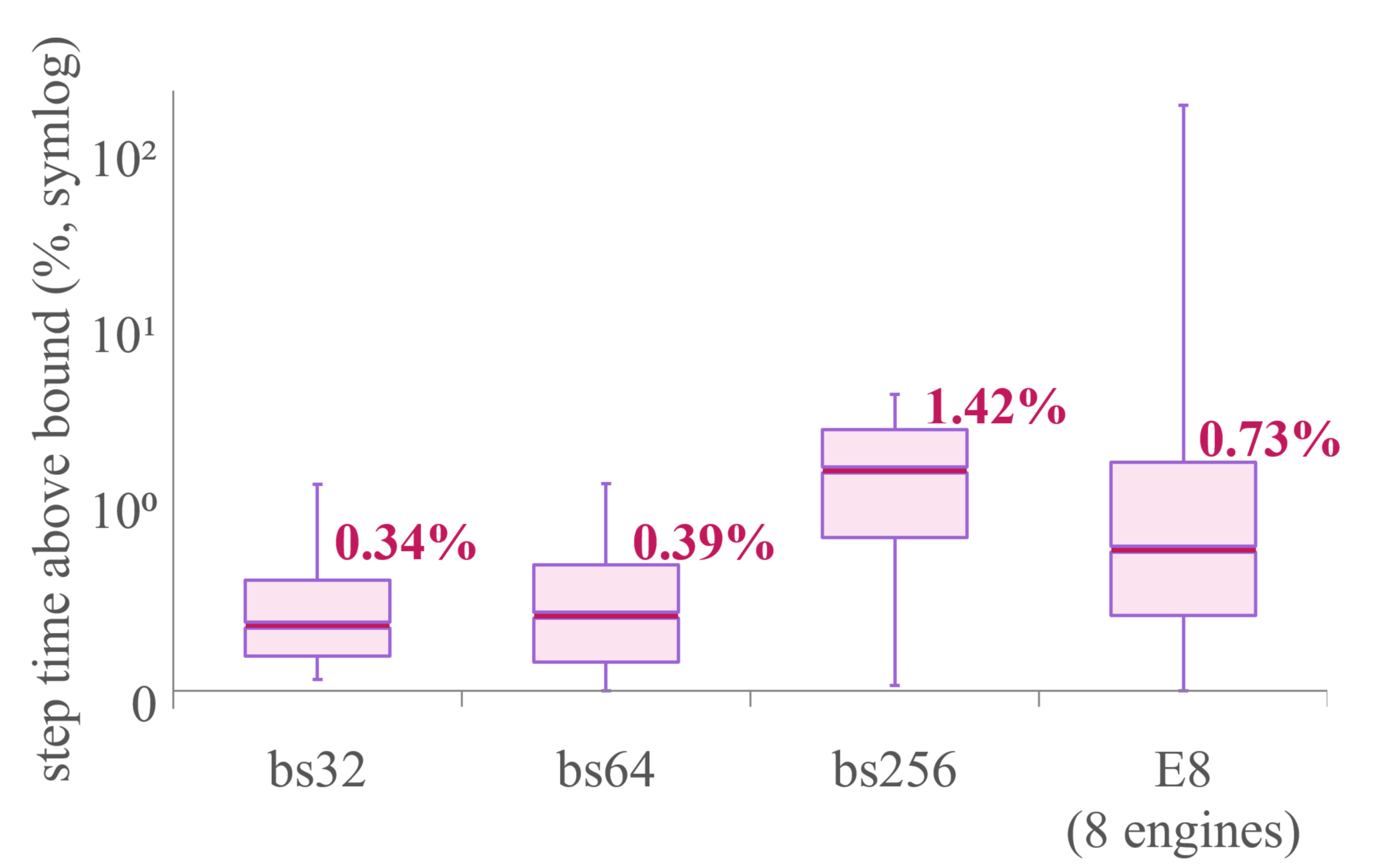}
\par\scriptsize (c) Distance above makespan bound
\end{minipage}\hfill
\begin{minipage}[t]{0.235\textwidth}
\centering
\includegraphics[width=\linewidth]{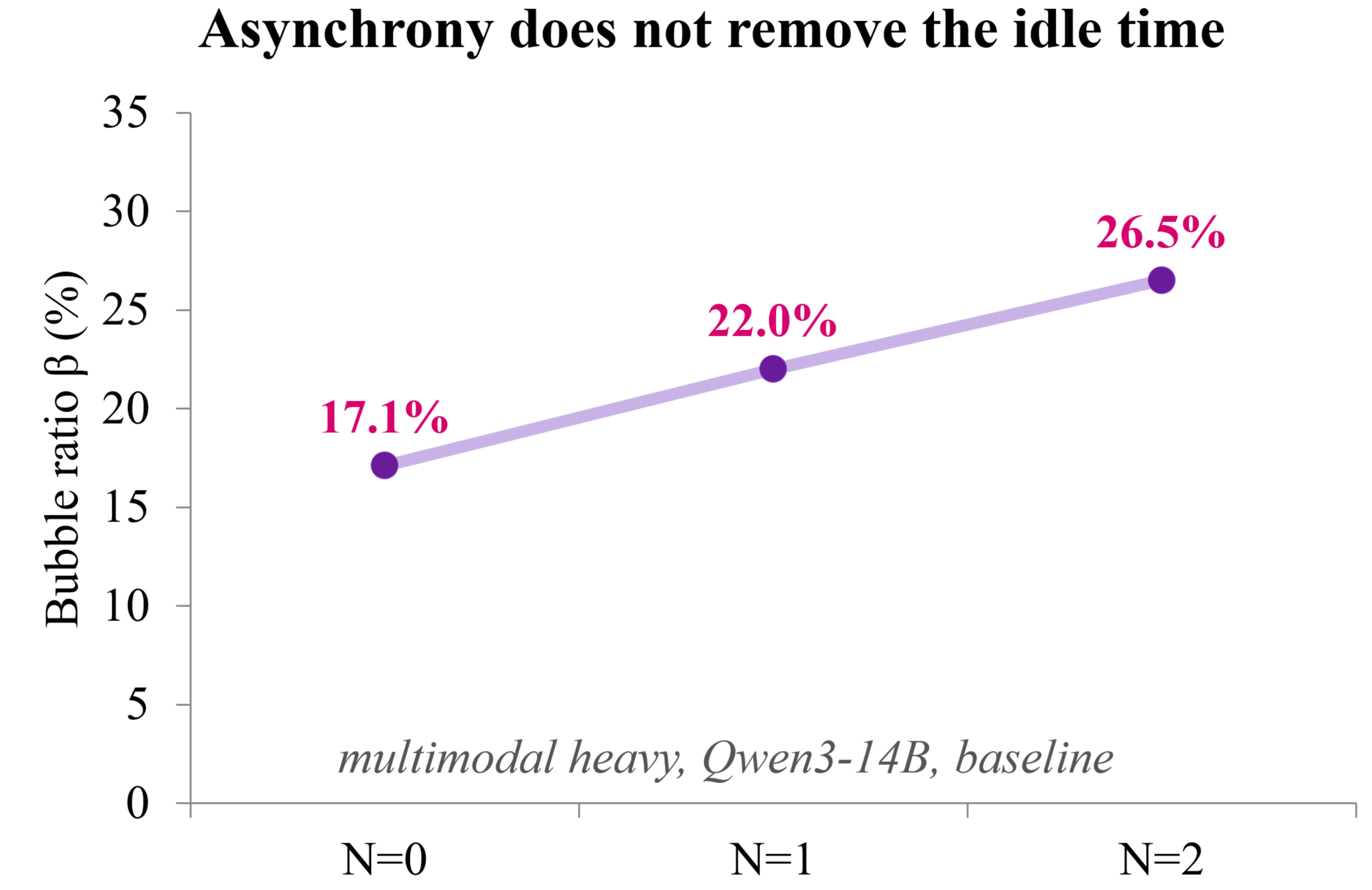}
\par\scriptsize (d) Bubble ratio vs.\ staleness $N$
\end{minipage}
\caption{\scriptsize Motivation measurements. (a)~Rollout share of the step: top, multimodal-heavy Qwen3-14B vs.\ 32B; bottom, Qwen3-14B AMC12 vs.\ Multitask; share rises 50\% to 76\% with response length. (b)~Median (p50) and p99 response lengths; all but capped AMC12 have a long tail. (c)~Distance above the makespan bound, $100 \times (\mathrm{span}/\mathrm{span}_{\min} - 1)$, over the batch/engine sweep (multimodal-heavy, $N{=}0$); boxes median/IQR, points steps. (d)~Bubble ratio vs.\ staleness $N$ (multimodal-heavy, baseline); asynchrony does not reduce engine idle time.}
\label{fig:findings}
\end{figure*}

\textbf{Insight 2: a few responses are much longer than the rest, so engines finish at very different times.}
Table~\ref{tab:workloads} and Fig.~\ref{fig:findings}(b) show the response-length spread across our workloads. On the multimodal-heavy workload, the median response length is 77 tokens, while the 99th-percentile response length is 2\,664 tokens---about 37 times the median. The multimodal-normal workload and multitask show similar patterns, with p99 response lengths 11 to 18 times their medians, while CodeAgent (ep.) reaches 3.9 times. AMC12 is the exception: its 8K-token generation cap clips the upper tail, keeping p99 at only 1.6 times the median (whereas a larger cap would expose a wider spread). This variation means that a step can contain mostly short responses alongside a few much longer ones. An engine processing a long response continues decoding after engines handling shorter responses have finished and become idle. The resulting bubble ratio in Table~\ref{tab:workloads} ranges from 0.17 to 0.33 across the 14B baseline workloads and reaches 0.57 on the 32B heavy run. The bubble ratio does not depend on the p99/p50 ratio alone, however: multimodal-normal has a smaller length spread than multimodal-heavy but a larger bubble ratio, as workloads also differ in total token count, prompt length, and the number of requests still running near the end of a step. Length variation creates the opportunity for imbalance but does not fully determine how much engine time is lost, motivating the per-step decomposition in Sec.~\ref{sec:formulation}.

\textbf{Insight 3: the longest request determines how long a step takes.}
A rollout step cannot finish until its longest request finishes, and it cannot finish faster than the total work divided by the number of engines. Both are lower bounds on the step time, so the operative bound is their maximum, $\mathrm{span}_{\min}=\max(\bar{b},L)$ (Sec.~\ref{sec:formulation}); the maximum is the tighter, binding bound. Table~\ref{tab:bound} shows that, in every workload, the longest request $L$ takes longer than the average-work term $\bar{b}$, so $L$ is the binding term. Fig.~\ref{fig:findings}(c) shows how closely the measured step time approaches this bound across the batch-size sweep of Sec.~\ref{sec:eval-sensitivity} and the eight-engine configuration of Sec.~\ref{sec:eval-headline}. The median step finishes only 0.3--1.4\% above the bound across all tested configurations, including just 0.7\% with eight engines. The mean for the eight-engine configuration is much higher, at 32\%, because a few steps have unusually large gaps, reaching 165\%. This insight limits what load balancing can do. Moving requests between engines can fill idle time, and Sec.~\ref{sec:eval-levers} shows that it does. But if the step is already at its bound, filling idle time does not make the step shorter. Only two things can: making the longest request itself decode faster, or not letting it run to the end.

\textbf{Insight 4: asynchronous training does not remove the idle time.}
An RL framework can let the trainer start the next update before every response of the current step is finished, using data that is up to $N$ steps old. This removes the wait from the trainer, but it does not change the rollout itself, which is why the bubbles survive. The bubble ratio is measured \emph{inside} each rollout window: it counts engine time that sits idle while that window is open, regardless of what the trainer is doing meanwhile. In fact it grows: to keep the trainer fed, higher staleness admits more concurrent rollouts, and those extra requests pile onto engines under the same round-robin pinning, deepening the queues on loaded engines while drained engines still cannot take over their work. On the multimodal-heavy workload, the baseline bubble ratio is 0.171 at $N{=}0$, 0.220 at $N{=}1$, and 0.265 at $N{=}2$ (Fig.~\ref{fig:findings}(d)). A fix therefore has to act inside rollout, and it has to work when the trainer is not waiting at a barrier.

\textbf{What these measurements require of a design.} Three requirements follow. First, the system must act \emph{during} decode: the imbalance in Insight~2 appears after requests start, and Insight~3 shows dispatch-time placement cannot shorten a step already at its bound (Sec.~\ref{sec:eval-baselines} tests this directly: a dispatcher that knows every output length in advance finishes within 0.2\% of the stock dispatcher). Second, the decision to move a request must be based on remaining work and move cost, not request counts: engines hold similar numbers of requests while doing very different amounts of work, and moving a request requires the destination to rebuild its state (Sec.~\ref{sec:mechanism}). Third, because the longest request sets the step time, the system needs a way to act on that request itself---indirectly via migration, by moving other requests off its engine so it decodes faster, or directly by stopping it early, which changes the training data and so needs a rule for when that is safe. Secs.~\ref{sec:mechanism}--\ref{sec:termination} build these three parts.

%% file: sections/4.System-overview.tex
\section{\sys{} System Overview}
\label{sec:overview}

\begin{figure}[t]
  \centering
  \includegraphics[width=0.92\columnwidth]{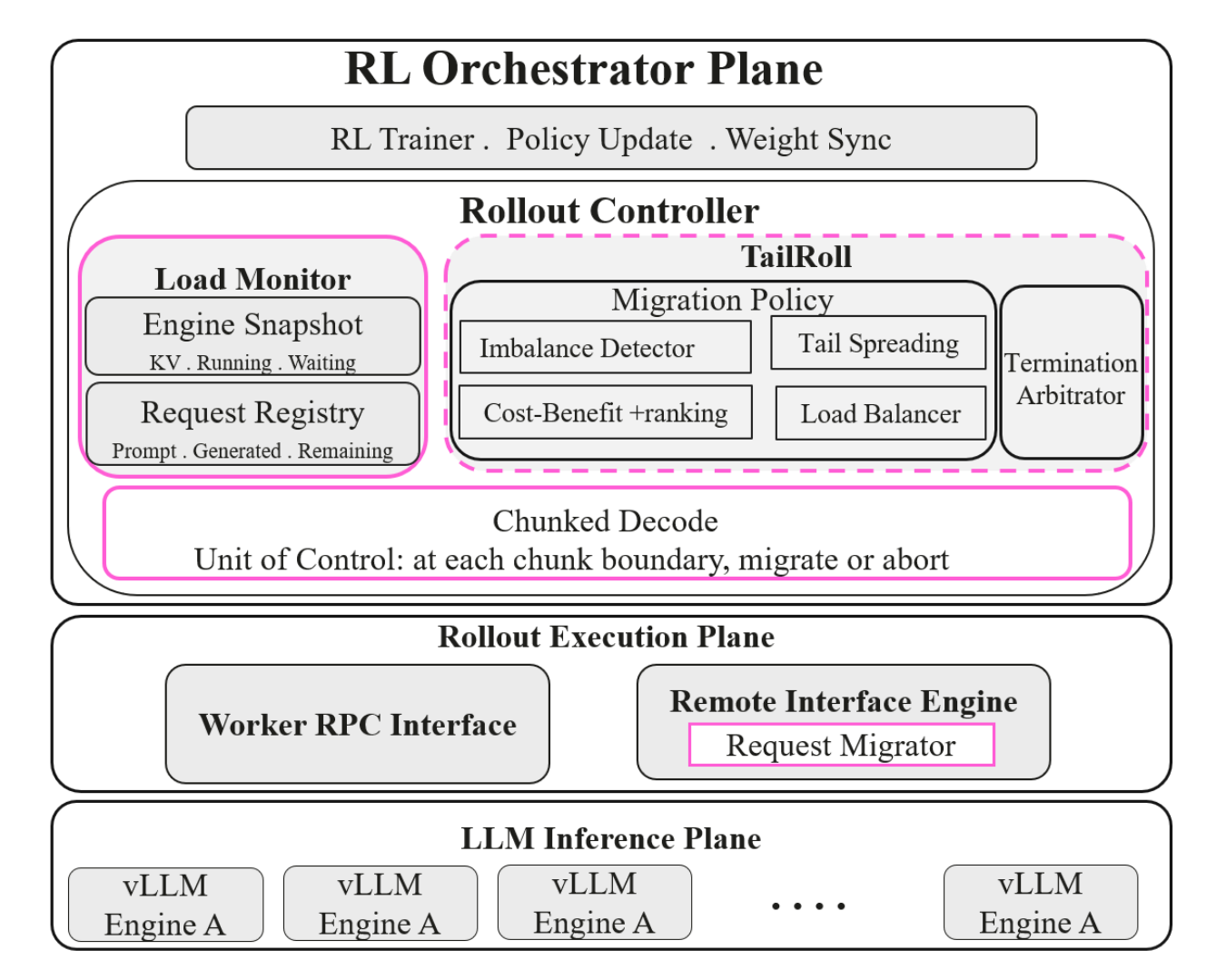}
  \caption{\scriptsize \sys{} architecture. \sys{} adds the load monitor and decision logic to the \emph{Rollout Controller} and the request migrator and abort endpoint to the \emph{Rollout Execution Plane}; the \emph{RL Orchestrator Plane} (\areal{}) and \emph{LLM Inference Plane} (vLLM) are unchanged. Chunked decode is the migration/abort boundary.}
  \label{fig:arch}
\end{figure}

\sys{} adds a lightweight control layer to \areal{}'s rollout path without modifying the model, loss, inference kernel, or transport protocol. Fig.~\ref{fig:arch} shows the four-plane architecture. The two inner planes---\emph{Rollout Controller} and \emph{Rollout Execution}---are where \sys{} acts; the \areal{} orchestrator and vLLM inference planes are unchanged.

\sys{} operates as an \emph{observe--decide--execute} loop. A \emph{load monitor} polls each engine every $\tau$ seconds, recording KV-cache utilization, request counts, and per-request progress in a registry. Each poll produces one snapshot that a \emph{migration policy} uses to decide whether to migrate a request or terminate a straggler; decisions are recorded as markers rather than applied immediately. \emph{Chunked decode} provides the execution boundary: each request re-enters the controller once per chunk of $C$ tokens, picks up any pending marker, and either hands off to a target engine via re-prefill or invokes the abort endpoint. The key design point is that \emph{observation and decision run on a wall-clock period} ($\tau{=}3$\,s) \emph{while execution runs on a token-count period} ($C$ tokens per chunk). Decoupling the two clocks means no decision interrupts a running kernel, no engine is probed on the request path, and reaction time is bounded to at most one chunk. Secs.~\ref{sec:formulation}--\ref{sec:termination} give the formulation, mechanism (Sec.~\ref{sec:mechanism}), policy, and termination logic in full.

%% file: sections/5.TailRoll-Design.tex
\section{\sys{} Design and Problem Formulation}
\label{sec:design}

\subsection{Problem Formulation}
\label{sec:formulation}

\textbf{Waste and bubble ratio.}
Consider one rollout step served by $E$ inference engines. For each engine $e$, let $\mathrm{busy}_e$ denote the total time during which the engine is actively processing at least one request, and let $\mathrm{span}$ denote the wall-clock duration of the rollout step, measured from the start of the first request to the completion of the last request. We define the wasted engine-time ($W$) and the \emph{bubble ratio} ($\beta$) as
\begin{equation}
  W = E\cdot\mathrm{span} - \sum_e \mathrm{busy}_e,
  \quad
  \beta = 1 - \frac{\sum_e \mathrm{busy}_e}{E\cdot \mathrm{span}}.
  \label{eq:bubble}
\end{equation}

\textbf{Imbalance versus straggler waste.}
A placement mechanism can move a request between engines but cannot parallelize its sequential token generation. Let $\bar{b}$ be the average engine busy time and $L$ be the longest execution time of any single request on an engine in the step:
\begin{equation}
  \begin{aligned}
    \bar{b} &= \frac{1}{E}\sum_e \mathrm{busy}_e,
    & L &= \max_r(\text{execution time of request } r),\\
    \mathrm{span}_{\min} &= \max(\bar b,\, L).
  \end{aligned}
  \label{eq:terms}
\end{equation}
Here, $\mathrm{span}_{\min}$ represents the makespan lower bound: a rollout step cannot finish faster than either the average work per engine ($\bar{b}$) or the longest individual request execution ($L$). Using $\mathrm{span}_{\min}$, the total waste splits exactly into two components:
\begin{equation}
  W = \underbrace{E\big(\mathrm{span}-\mathrm{span}_{\min}\big)}_{W_{\mathrm{imb}}}
     + \underbrace{E\big(\mathrm{span}_{\min}-\bar b\big)}_{W_{\mathrm{strag}}}.
  \label{eq:split}
\end{equation}
$W_{\mathrm{imb}}$ is the \emph{imbalance waste} that better load balancing can recover---it arises when requests queue or slow each other down on a busy engine while other engines sit idle. $W_{\mathrm{strag}}$ is the \emph{straggler waste} caused by a single long-running straggler request that keeps the step open even after all other engines have drained. In \sys{}, in-flight migration targets $W_{\mathrm{imb}}$, while sufficiency-gated termination targets $W_{\mathrm{strag}}$.

\textbf{Objective.}
In each RL iteration, the end-to-end step latency is $T_{\mathrm{step}} = \mathrm{span} + T_{\mathrm{train}}$, and the effective rollout throughput across the cluster is $N_{\mathrm{tok}}/\mathrm{span}$, where $N_{\mathrm{tok}}$ is the total number of generated tokens. Because $T_{\mathrm{train}}$ depends only on the collected batch, minimizing the rollout duration $\mathrm{span}$ (equivalently, minimizing total waste $W = W_{\mathrm{imb}} + W_{\mathrm{strag}}$) directly minimizes end-to-end training time and maximizes cluster throughput:
\begin{equation}
\begin{aligned}
  &\min \textstyle\sum_{\text{steps}} T_{\mathrm{step}}
  \;\;\Longleftrightarrow\;\;
  \min \textstyle\sum_{\text{steps}} \big(W_{\mathrm{imb}} + W_{\mathrm{strag}}\big) \\
  &\quad\text{s.t.}\quad \mathbb{E}[\text{reward}]\ \text{unchanged}.
\end{aligned}
\label{eq:objective}
\end{equation}
Migration reduces $W_{\mathrm{imb}}$ while preserving the generated token sequence, whenever its re-prefill overhead is smaller than the waste it recovers. Termination reduces $W_{\mathrm{strag}}$ by truncating extreme stragglers and preserves policy reward through a group-sufficiency condition on GRPO groups (Sec.~\ref{sec:termination}).

\subsection{In-Flight Migration Mechanism}
\label{sec:mechanism}

\textbf{Chunked decode.}
\sys{} divides each request's generation into \emph{chunks} of $C$ tokens (default $C{=}512$). The generation loop issues HTTP calls with \texttt{max\_new\_tokens} capped at $C$, accumulates the returned tokens, and re-issues until a natural stop or the budget is reached. Because vLLM prefix caching retains the request's KV blocks between chunks on the same engine, chunking costs one HTTP round trip per chunk and no recomputation---under 0.5\% overhead in our measurements. At each chunk boundary, the loop checks (in order): (1) whether a weight update is in progress; (2) whether a termination marker is pending; (3) whether a migration marker has been set. These three checks share one control point, so no migration is ever issued while an engine is loading fresh weights.

\textbf{Stateless handoff.}
When the policy selects request $r$ on source $s$ for target $t$, the controller records a marker and returns immediately---$s$ keeps decoding. At $r$'s next chunk boundary: (1) the loop pops the marker; if $r$ is within 512 tokens of its budget, the move is cancelled (finishing locally is cheaper); (2) the loop rebinds $r$ to $t$ in the request registry; (3) the next chunk is issued to $t$ with \texttt{input\_ids} = prompt $\Vert$ all generated tokens, and the target performs one forward pass over the entire known sequence in parallel (not chunked prefill, which would divide it into smaller batches), rebuilding the KV cache, and continues auto-regressive decode from where $s$ stopped; and (4) on $s$, KV blocks are released by the engine's normal completion path.

\textbf{Migration cost.}
The cost of migrating a request with prompt length $p$ and $g$ generated tokens is the re-prefill time expressed in decode-token-equivalents:
\begin{equation}
  \mathrm{cost}(p, g) = \frac{p + g}{\sigma},
  \label{eq:cost}
\end{equation}
where $\sigma$ is the prefill-to-decode throughput ratio. On our hardware, $\sigma \approx 10$: prefill runs roughly ten times faster per token than decode because it is fully parallelizable. The policy charges this cost against every candidate move (Sec.~\ref{sec:selection}).

\textbf{Telemetry.}
A background load monitor wakes every $\tau$ seconds (default 3\,s) and queries each engine's Prometheus counters (KV-cache utilization, running and waiting counts); all values in a poll are frozen as one snapshot.

\subsection{Migration Policy}
\label{sec:policy}

At each poll, \sys{} applies three mechanisms in order (Algorithm~\ref{alg:poll}): (1) tail-aware load balancing, (2) termination check, (3) general migration. This order gives simpler 
interventions priority. All thresholds are dimensionless ratios, so no workload-specific tuning is required.
\subsubsection{Imbalance Detection}
\label{sec:detect}

We detect imbalance using a \emph{freeness} score $f_e = \frac{1 - \mathrm{KV}_e}{\max(1,\; \mathrm{running}_e)}$ that combines running-request counts with KV-cache usage. Let $s=\arg\min_e f_e$ (least free) and $t=\arg\max_e f_e$ (most free). We declare an imbalance when $s$ has at least two running requests and $f_t/f_s \ge \rho$ ($\rho{=}3$). A \emph{late-tail override} handles the end of a step: if one engine holds only one running request while at least two other engines are idle, we declare an imbalance even when the normal threshold is not met. This catches cases where a single remaining request keeps the step open while most of the cluster sits idle. To avoid reacting to transient fluctuations, the imbalance must persist for two consecutive polls. After a migration, the source engine enters a cooldown before it can be selected again.
\subsubsection{Tail-Aware Load Balancing}
\label{sec:spread}

The freeness ratio becomes less reliable when few requests remain. As long as one engine holds multiple requests while another engine sits idle, however, work can be redistributed:
\begin{equation}
  \exists\,e:\ \mathrm{running}_e \ge 2 \ \wedge\ \exists\,e':\ \mathrm{running}_{e'} = 0.
  \label{eq:tailbalance}
\end{equation}
Whenever Eq.~\eqref{eq:tailbalance} holds, \sys{} bypasses the freeness ratio and the two-poll requirement. It moves requests one at a time from the most-loaded engine to an idle engine until every running request occupies its own engine or no idle engine remains. This mechanism is inherently group-aware: \areal{} co-locates the $n$ sibling samples of a prompt on one engine to reuse the prompt's KV cache. Requests stacked on one engine near the tail are therefore usually siblings from the same GRPO group, and redistributing them lets siblings decode in parallel rather than sequentially. Because siblings typically have similar lengths (they share a prompt and the same stopping rule), they often finish near the same time, so spreading them across idle engines allows the entire group to complete faster.

\subsubsection{Request Selection}
\label{sec:selection}

For each source--target pair, \sys{} considers requests with at least 128 tokens remaining and a registry entry refreshed within 30\,s. A request that has already migrated once is never moved again.
\textbf{Cost-benefit filter.}
Let $p$ be the prompt length, $g$ the tokens generated, $m$ the tokens remaining, and $q = (\mathrm{running}_s{+}1)/(\mathrm{running}_t{+}1)$ the queue-depth ratio. The expected benefit is decoding $m$ tokens on a less crowded engine, estimated as $m\,(q{-}1)$ decode-token-equivalents. We migrate only when:
\begin{equation}
  m \cdot \max(q-1,\,0.1) > \frac{p+g}{\sigma}.
  \label{eq:cb}
\end{equation}
Requests that have generated more than 80\% of their budget are never moved, because their remaining decode work is unlikely to repay the re-prefill cost and they are close to finishing naturally. Requests with $g{=}0$ bypass this filter since they have no KV state to rebuild.

\textbf{Ranking.}
After filtering, candidates are ranked by $\mathrm{score}(r) = \frac{g \cdot m}{(p+M)^2}$, where $M$ is the request's maximum new tokens. The numerator favors requests with accumulated state and substantial remaining work; the denominator normalizes by footprint. Requests with $g{=}0$ are ranked by $m/M$ and moved first---they cost no re-prefill.

\subsubsection{Volume Control}
\label{sec:volume}

The number of requests moved is bounded by three limits, $n = \min(n_{\mathrm{count}},\, n_{\mathrm{kv}},\, n_{\mathrm{room}})$ with $n \le \mathrm{running}_s - 1$. $n_{\mathrm{count}} = \lfloor(\mathrm{running}_s - \mathrm{running}_t')/2\rfloor$ moves only enough to bring counts halfway toward equal. $n_{\mathrm{kv}}$ and $n_{\mathrm{room}}$ similarly limit moves based on KV-cache differences and target capacity. At most 8 requests move in one batch to avoid stalling the target's decoding. Each moved request reserves target headroom for 30\,s.

\subsection{Sufficiency-Gated Termination}
\label{sec:termination}

Once every running request occupies its own engine, the remaining step time is determined by the longest request---the straggler waste $W_{\mathrm{strag}}$ in Eq.~\eqref{eq:split}.

\textbf{Makespan trigger.}
Termination becomes eligible when the number of running requests falls below half the step's peak concurrency. At this point, the step is in its late tail, and the longest request is likely determining the final completion time. The arbitrator considers only the top 5\% of running requests by remaining work. The leading request must have at least $1.5\times$ the remaining work of the first request outside this group, distinguishing true stragglers from requests close in completion time.

\textbf{Group sufficiency.}
Even when a request meets the makespan test, it is terminated only if its GRPO group can spare it. Let $k(r)$ be the number of $r$'s siblings (out of $n$) that have already finished. All three conditions must hold: (1) \emph{Sufficiency:} $k(r) \ge k_{\min}$ ($k_{\min}{=}3$ with $n{=}4$), so the group has enough finished siblings to form its advantage estimate; (2) \emph{Minimum length:} $r$ has generated at least 128 tokens, so the partial trajectory is not degenerate; (3) \emph{Truncation cap:} terminations this step stay below 5\% of the step's peak concurrency, bounding any shift in the training distribution.

\textbf{Execution.}
Termination uses \texttt{abort\_request} endpoint. The request stops decoding and its KV blocks are freed. The tokens generated so far are returned as a completed-but-truncated trajectory, with the loss masked past the last token. The partial trajectory is retained, not discarded.

\begin{algorithm}[t]
\scriptsize
\caption{\scriptsize \sys{}: tail-aware rollout optimization (per poll)}
\label{alg:poll}
\vspace{-3pt}
\begin{algorithmic}[1]
\setlength{\itemsep}{0pt}
\setlength{\parskip}{0pt}
\linespread{0.92}\selectfont

\STATE $T \gets$ \texttt{get\_server\_loads}() for all engines
\hfill$\triangleright$ one snapshot
\IF{$\exists e:\mathrm{running}_e \ge 2$ \textbf{and}
    $\exists e':\mathrm{running}_{e'} = 0$}
  \WHILE{such $e, e'$ exist}
    \STATE $r \gets$ best candidate on
    $\arg\max_e \mathrm{running}_e$ passing Eq.~\eqref{eq:cb}
    \STATE \texttt{request\_migration}$(r, e')$
    \hfill$\triangleright$ split co-located groups
  \ENDWHILE
  \IF{any request moved}
    \STATE \textbf{return}
  \ENDIF
\ENDIF
\IF{$\sum_e \mathrm{running}_e \le
    \tfrac{1}{2}\cdot\mathrm{peak}$}
  \FOR{each $r$ in the top 5\% by remaining work, with remaining work
       $\ge 1.5\times$ pivot, $k(r) \ge k_{\min}$,
       $g_r \ge 128$, and termination cap not hit}
    \STATE \texttt{abort}$(r)$; keep masked partial
  \ENDFOR
  \IF{any request terminated \textbf{or} no engine is stacked}
    \STATE \textbf{return}
  \ENDIF
\ENDIF
\STATE $f_e \gets (1-\mathrm{KV}_e)/
\max(1,\mathrm{running}_e)$ for all $e$
\IF{not imbalanced($f$) for 2 consecutive polls}
  \STATE \textbf{return}
\ENDIF
\FOR{each pair $(s,t)$ in the greedy pairing by $f$,
     with cooldown expired and gap $\ge 4$}
  \STATE $\mathcal{C} \gets$ candidates on $s$ passing Eq.~\eqref{eq:cb}
  \STATE $n \gets \min(n_{\mathrm{count}},
  n_{\mathrm{kv}}, n_{\mathrm{room}}, 8)$
  \FOR{each of the top-$n$ candidates $r \in \mathcal{C}$ by score}
    \STATE \texttt{request\_migration}$(r,t)$;
    reserve slot on $t$ (30\,s)
  \ENDFOR
  \STATE start cooldown on $s$; schedule verification re-poll
\ENDFOR

\end{algorithmic}
\vspace{-4pt}
\end{algorithm}

%% file: sections/6.Implementation.tex
\section{Implementation and Experimental Setup}
\label{sec:impl}

\textbf{Implementation.} \sys{} is implemented in about 3,700 lines of Python on top of \areal{}, chiefly in \texttt{RolloutController} and \texttt{RemoteInfEngine}, plus a small extension of \areal{}'s vLLM server wrapper that exports per-engine load metrics and an abort endpoint. It runs unchanged on vLLM over Ascend NPUs and NVIDIA GPUs through the same backend interface. All three mechanisms share one control point at the request's next chunk boundary; termination reuses the existing \texttt{abort\_request} endpoint. Policy knobs (chunk size, imbalance threshold $\rho$, detection signal, cost-benefit gate, $k_{\min}$) are set per run through environment variables, with one switch per mechanism for the ablations below. Baseline and \sys{} runs share a single build: the baseline keeps the load monitor running with every mechanism disabled, so poll overhead is paid by both arms and runs differ only by configuration. Every job writes a sidecar recording its configuration, environment, and git revision, so each table cell below is traceable to a run.

\subsection{Experimental Setup}
\label{sec:eval-setup}

\textbf{Testbed.} Two Ascend nodes with 16 910C NPUs each (64\,GB HBM per NPU); a run uses one or both nodes depending on its configuration. By default, eight NPUs run rollout as 4 vLLM engines (TP=2 per engine) and eight run training as 8 FSDP ranks.

\textbf{Controlling for engine-speed skew.}\label{sec:impl-health} Every result below is a difference between two runs, so it can be attributed to a scheduling policy only if the engines in both runs decode at comparable speed. Before the first step, the controller sends one identical 48-token request to every engine in parallel and logs, as a per-step metric, the \emph{engine skew}: the ratio between the highest and lowest measured per-engine decode rates within a launch. We treat a launch as comparable if its skew is at most $1.3\times$, and screen reported comparisons against this threshold. Across the reported runs, skew ranges from $1.01\times$ to $1.33\times$ (plus $1.60\times$ on the 16K AMC12 full arm). At 8K, the full-\sys{} arm of AMC12 ($1.33\times$) marginally exceeds the threshold; the other AMC12 arms stay within it ($1.08$--$1.21\times$) and show the same direction of effect (Sec.~\ref{sec:eval-realtask}).

\textbf{Models.} Qwen3-32B for the headline end-to-end result (Sec.~\ref{sec:eval-headline}) and Qwen3-14B for the mechanism and policy studies (Secs.~\ref{sec:eval-levers}--\ref{sec:eval-overhead}); bf16; vLLM with prefix caching, \texttt{max\_num\_seqs}=1024, and 80\% memory utilization.

\textbf{Algorithm.} GRPO via \areal{}'s decoupled PPO loss; $B{=}128$ prompts $\times$ $n{=}4$ samples; \texttt{max\_new\_tokens}=3072; up to 1024 concurrent rollouts; staleness $N{=}0$ (sync) and $N{=}1$ (async).

\textbf{Workloads.} (i)~\emph{Multimodal-normal} and \emph{multimodal-heavy}: response-length distributions replayed from the Azure multimodal LLM inference trace~\cite{azuretrace}. (ii)~\emph{Straggler pool}: manufactured distribution---short bulk plus a rare 0.4\% tail at 1400--3072 tokens. (iii)~\emph{CodeAgent}: a multi-turn code-agent trace replayed at turn level and episode level. (iv)~\emph{Multitask}: equal-weight mix of multimodal, CodeAgent, GSM8K-natural, and AMC12-natural lengths. (v)~\emph{AMC12}: native workload---real competition problems with the model's own reasoning lengths (thinking enabled, evaluated at 8K and 16K token budgets) and exact-match reward. (vi)~\emph{Multimodal (live reward)}: uses the multimodal-normal trace with natural stopping, and a live training reward. This is reported as a 50-step matched pair and a 300-step training run. Table~\ref{tab:workloads} summarizes length statistics.

\textbf{Baselines.}
\begin{itemize}
    \item \textbf{Vanilla \areal{}}: The unmodified \areal{} rollout path with its stock round-robin dispatcher, used as the reference point against which every mechanism is measured.
    \item \textbf{Least-loaded dispatch}: A count-based dispatch policy we implemented on top of \areal{}, which routes each arriving request to the engine with the fewest running requests.
    \item \textbf{Oracle-work dispatch}: A length-aware dispatcher we built as an optimistic reference for dispatch-time placement; it is given each request's true output length and the true remaining work of every in-flight request, which is strictly more information than any length predictor can supply (Sec.~\ref{sec:eval-baselines}).
    \item \textbf{Seer-style detection}: An adaptation of Seer's trigger~\cite{qin2026seer} to our setting, in which we drive the \sys{} migration mechanism from Seer's running-request-spread signal rather than our freeness signal, isolating the effect of the detection signal alone.
\end{itemize}

\textbf{\sys{} configurations.} To isolate the contribution of each component, we evaluate \sys{} by enabling its three mechanisms cumulatively: (1)~\textbf{Migration}, which enables freeness-triggered in-flight request migration with cost-benefit filtering during bulk execution; (2)~\textbf{+Tail-aware balancing}, which additionally redistributes co-located tail requests onto idle engines once bulk queues drain (Eq.~\eqref{eq:tailbalance}); and (3)~\textbf{+Termination}, the full \sys{} system, which adds sufficiency-gated early termination ($k_{\min}{=}3$) for residual stragglers.

\textbf{Metrics.} \emph{Rollout s/step} is the headline number; \emph{bubble ratio} $\beta$ (Eq.~\eqref{eq:bubble}), split into $W_{\mathrm{imb}}$ and $W_{\mathrm{strag}}$ (Eq.~\eqref{eq:split}), is the proxy. We also report \emph{$\mathrm{span}/\mathrm{span}_{\min}$} (distance from the makespan bound, Eq.~\eqref{eq:terms}), cross-engine \emph{KV-cache std and peak occupancy}, \emph{worst-engine stall time}, per-request latencies (\emph{TTFT}, \emph{TPOT}, and \emph{end-to-end request latency}), \emph{tokens/s}, mean \emph{reward}, and \emph{truncation rate}. All figures are means over matched steps, dropping each run's first two steps (weight load and first-token warm-up).

%% file: sections/7.Evaluation.tex
\section{Evaluation}
\label{sec:eval}

Our evaluation addresses six questions: end-to-end waste reduction across workloads, model scales, and engine counts (\textbf{Q1}, Sec.~\ref{sec:eval-headline}); gains on real RL tasks with live rewards (\textbf{Q2}, Sec.~\ref{sec:eval-realtask}); migration's improvement over dispatch-time placement (\textbf{Q3}, Secs.~\ref{sec:eval-bound}--\ref{sec:eval-baselines}); the contribution of the freeness signal, cost-benefit gate, and tail-aware balancing (\textbf{Q4}, Sec.~\ref{sec:eval-policy}); sensitivity to staleness, group size, and batch size (\textbf{Q5}, Sec.~\ref{sec:eval-sensitivity}); and mechanism overhead (\textbf{Q6}, Sec.~\ref{sec:eval-overhead}). Table~\ref{tab:summary} summarizes the headline results below. Setup, baselines, and metrics are given in Sec.~\ref{sec:impl}; the engine-skew check applied to every comparison is described in Sec.~\ref{sec:impl-health}.
\begin{table}[t]
\centering
\caption{\scriptsize Rollout time and bubble ratio, full \sys{} ($N{=}0$). $^\dagger$Stretched multimodal-heavy tail, migration only.}
\label{tab:summary}
\small
\setlength{\tabcolsep}{3.5pt}
\resizebox{0.92\columnwidth}{!}{%
\begin{tabular}{ll|rrr|rrr}
\toprule
 & & \multicolumn{3}{c|}{Rollout time (s/step)} & \multicolumn{3}{c}{Bubble ratio $\beta$} \\
Workload & Scale & Baseline & \sys{} & Change & Baseline & \sys{} & Change \\
\midrule
\textbf{AMC12 (16K, natural)} & 14B & 1946.3 & \textbf{1287.1} & \textbf{$-33.9\%$} & 0.208 & 0.055 & $-73.7\%$ \\
\textbf{Multimodal (live reward)} & 14B & 15.9 & \textbf{10.5} & \textbf{$-33.8\%$} & 0.191 & 0.088 & $-53.8\%$ \\
\textbf{AMC12 (8K, natural)} & 14B & 644.7 & \textbf{492.8} & \textbf{$-23.6\%$} & 0.175 & 0.091 & $-48.0\%$ \\
CodeAgent (episode) & 14B & 541.0 & 519.5 & $-4.0\%$ & 0.278 & 0.123 & $-55.7\%$ \\
Multitask & 14B & 237.6 & 230.1 & $-3.2\%$ & 0.249 & 0.145 & $-41.9\%$ \\
Multimodal ($p99/p50{=}80$)$^\dagger$ & 14B & 202.3 & 196.0 & $-3.1\%$ & 0.237 & 0.099 & $-58.1\%$ \\
Multimodal-heavy & 14B & 90.0 & 90.2 & $+0.2\%$ & 0.171 & 0.079 & $-53.5\%$ \\
Multimodal-normal & 14B & 58.3 & 58.8 & $+0.9\%$ & 0.312 & 0.088 & $-71.9\%$ \\
\midrule
Multimodal-heavy & 32B & 139.2 & 140.1 & $+0.6\%$ & 0.573 & 0.295 & $-48.5\%$ \\
Multimodal-heavy ($E{=}8$) & 14B & 86.2 & 85.6 & $-0.7\%$ & 0.372 & 0.295 & $-20.6\%$ \\
\bottomrule
\end{tabular}}
\end{table}
\subsection{End-to-End Waste Reduction}
\label{sec:eval-headline}
\label{sec:eval-levers}

\begin{figure}[t]
  \centering
  \includegraphics[width=0.85\columnwidth]{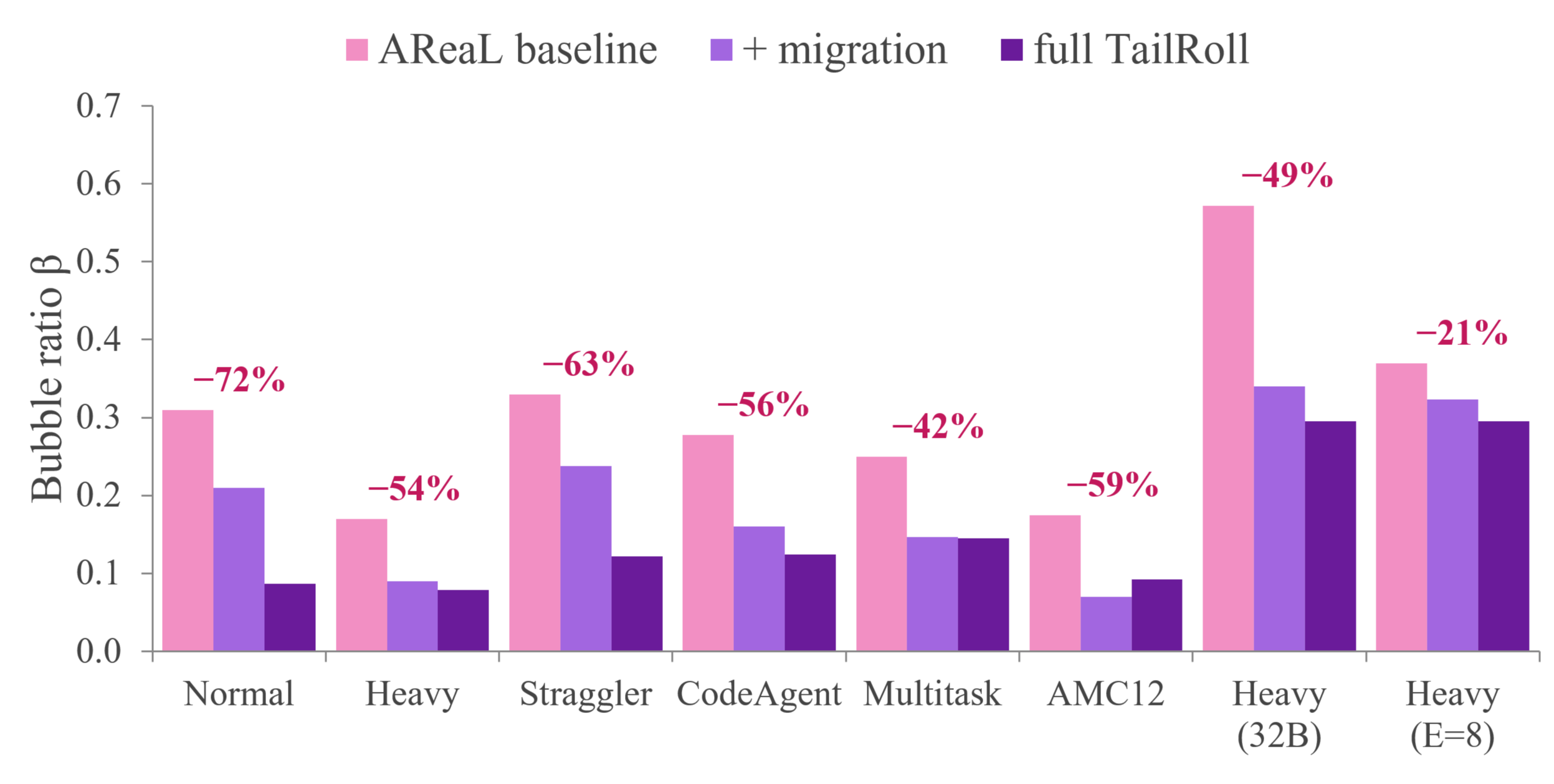}
  \caption{\scriptsize Bubble ratio across workloads/scales ($N{=}0$, Qwen3-14B unless marked). \sys{} cuts waste by 21--74\%.}
  \label{fig:eval-bubble}
\end{figure}
\begin{table}[t]
\centering
\caption{\scriptsize Lever study ($N{=}0$, Qwen3-14B), cumulative arms. Replayed traces: output tokens identical for migration/balancing, $\le$0.015\% different with termination.}
\label{tab:levers}
\small
\setlength{\tabcolsep}{3.5pt}
\resizebox{0.92\columnwidth}{!}{%
\begin{tabular}{l|l|r|r}
\toprule
Workload & Arm & Rollout (s/step) & $\beta$ \\
\midrule
\multirow{4}{*}{Multimodal-normal} & \areal{} baseline & 58.3 & 0.312 \\
 & + migration        & 58.6 & 0.211 \\
 & + tail-aware balancing   & 58.4 & 0.076 \\
 & full \sys{} & 58.8 & 0.088 \\
\midrule
\multirow{4}{*}{Multimodal-heavy} & \areal{} baseline & 90.0 & 0.171 \\
 & + migration        & 89.4 & 0.092 \\
 & + tail-aware balancing   & 90.2 & 0.065 \\
 & full \sys{} & 90.2 & 0.079 \\
\midrule
\multirow{3}{*}{Straggler pool} & \areal{} baseline & 29.6 & 0.331 \\
 & + migration        & 29.2 & 0.237 \\
 & full \sys{}        & 29.8 & 0.122 \\
\midrule
\multirow{3}{*}{CodeAgent (episode)} & \areal{} baseline & 541.0 & 0.278 \\
 & + migration        & \textbf{517.7} & 0.162 \\
 & full \sys{}        & 519.5 & \textbf{0.123} \\
\midrule
\multirow{3}{*}{Multitask} & \areal{} baseline & 237.6 & 0.249 \\
 & + migration        & \textbf{227.4} & 0.148 \\
 & full \sys{}        & 230.1 & 0.145 \\
\midrule
\multirow{3}{*}{AMC12 (natural, real reward)} & \areal{} baseline & 644.7 & 0.175 \\
 & + migration        & 565.1 & \textbf{0.071} \\
 & full \sys{}        & \textbf{492.8} & 0.091 \\
\bottomrule
\end{tabular}}
\end{table}
Fig.~\ref{fig:eval-bubble} and Table~\ref{tab:levers} give the end-to-end result across six workloads at 14B, plus the 32B and $E{=}8$ scale settings.

\textbf{Bubble-ratio reduction across workloads.} Migration alone reduces the baseline bubble ratio by 28--60\% (lowest on straggler pool $0.331 \to 0.237$, highest on AMC12 $0.175 \to 0.071$). Tail-aware balancing provides further reductions where reported (normal: $0.211 \to 0.076$; heavy: $0.092 \to 0.065$). The full configuration increases $\beta$ slightly on some workloads relative to the preceding arm: termination removes decode work from a non-gating engine, shortening its busy time without shortening the step, which mechanically raises the idle fraction (Sec.~\ref{sec:eval-term} evaluates termination on span and reward). Full \sys{} achieves 42--72\% total bubble reduction across workloads. A second seed of the heavy trace confirms the pattern: baseline $88.6$\,s/$\beta{=}0.174$ vs.\ $85.7$\,s/$0.088$ for migration+balancing.

\textbf{Effect on rollout span.} Reducing $\beta$ does not always shorten the measured rollout span, and the two regimes in Table~\ref{tab:levers} illustrate why. On the episode-level CodeAgent replay the rollout step shortens by 4.3\% (541.0 $\to$ 517.7\,s), on the multitask trace by 4.3\% (237.6 $\to$ 227.4\,s), and on AMC12 by 12.3\% from migration alone (Sec.~\ref{sec:eval-realtask} analyzes AMC12 in depth). On the short-step multimodal traces, by contrast, span moves by at most $\sim$1\% while $\beta$ falls by half or more. The difference comes down to how much slack a step has above its makespan bound (Sec.~\ref{sec:eval-bound} formalizes this with the per-workload decomposition in Table~\ref{tab:bound}). On AMC12 and the CodeAgent episodes, concurrency stays high for most of the step, so moving requests off a loaded engine relieves decode contention on whichever request is currently gating the step; the longest job itself gets shorter, and the span falls with the bubble. On the short multimodal steps, the tail has already drained to a handful of requests, so the step is pinned by a single longest request that migration does not touch: the reclaimed engine-time cannot shorten that request, so it appears instead as throughput and latency headroom (heavy: $+$0.7\% tokens/s, p95 request latency 59.8 $\to$ 59.0\,s). In this drained-tail regime, only a decision about the straggler itself---termination---can move the critical path (Sec.~\ref{sec:eval-term}). We therefore always report rollout span beside $\beta$.
\begin{table*}[t]
\centering
\caption{\scriptsize Load balance, worst-engine stall, and serving gains ($N{=}0$, Qwen3-14B unless marked). KV std/max: cross-engine std / single-engine peak KV occupancy; worst stall: longest idle wait in a step.}
\label{tab:cluster-balance}
\scriptsize
\setlength{\tabcolsep}{3pt}
\renewcommand{\arraystretch}{1.0}

\begin{tabular}{@{}l|rr|r|rr|rr@{}}
\toprule
 & \multicolumn{2}{c|}{KV-cache std} 
 & Peak KV 
 & \multicolumn{2}{c|}{Worst engine stall (s)} 
 & \multicolumn{2}{c}{Tail latency / Tput} \\
Workload 
 & Base $\to$ \sys{} & $\Delta$ 
 & $\Delta$ 
 & Base $\to$ \sys{} & $\Delta$ 
 & Metric & $\Delta$ \\
\midrule
AMC12 (16K)       
& $0.131 \to 0.028$ & $-78.8\%$ & $-33.9\%$ 
& $803.2 \to 163.4$ & $-79.7\%$ 
& p95 TTFT / tok/s & $-70.7\% / {+}33.3\%$ \\

AMC12 (8K)        
& $0.100 \to 0.043$ & $-56.9\%$ & $-28.6\%$ 
& $220.3 \to 84.5$ & $-61.7\%$ 
& p95 TTFT / tok/s & $-88.4\% / {+}12.8\%$ \\

CodeAgent (ep.)   
& $0.037 \to 0.014$ & $-62.4\%$ & $-21.6\%$ 
& $300.2 \to 150.2$ & $-50.0\%$ 
& p95 vLLM / tok/s & $-10.2\% / {+}4.2\%$ \\

Multitask         
& $0.025 \to 0.008$ & $-68.0\%$ & $-27.2\%$ 
& $111.6 \to 66.2$ & $-40.7\%$ 
& p95 TTFT / tok/s & $-15.2\% / {+}4.1\%$ \\

Multimodal ($p99/p50{=}80$) 
& $0.021 \to 0.006$ & $-72.5\%$ & $-33.0\%$ 
& $104.3 \to 40.3$ & $-61.4\%$ 
& p99 E2E / tok/s & $-4.6\% / {+}3.0\%$ \\

Multimodal-heavy  
& $0.007 \to 0.003$ & $-58.3\%$ & $-21.2\%$ 
& $35.1 \to 15.2$ & $-56.7\%$ 
& mean TTFT / p95 E2E & $-4.1\% / {-}1.2\%$ \\

Multimodal-normal 
& $0.002 \to 0.001$ & $-44.4\%$ & $-13.2\%$ 
& $34.5 \to 10.8$ & $-68.8\%$ 
& p95 TPOT / mean TTFT & $-3.9\% / {-}3.2\%$ \\

Straggler pool    
& $0.002 \to 0.001$ & $-59.0\%$ & $-24.5\%$ 
& $22.2 \to 5.8$ & $-73.7\%$ 
& p99 E2E / tok/s & $-1.3\% / {+}1.5\%$ \\

\midrule
Heavy (32B)       
& $0.007 \to 0.002$ & $-68.6\%$ & $-26.4\%$ 
& --- & --- 
& p50 E2E / mean TTFT & $-7.4\% / {-}5.4\%$ \\

Heavy ($E{=}8$)   
& $0.007 \to 0.003$ & $-61.3\%$ & $-34.5\%$ 
& --- & --- 
& p95 E2E / tok/s & $-9.1\% / {+}3.7\%$ \\
\bottomrule
\end{tabular}
\end{table*}

\textbf{Cluster KV-cache balance, engine stalls, and request-level latency.} Even on workloads where a single longest trajectory bounds the step duration, \sys{} substantially improves cluster resource balance and individual request latency throughout the batch (Table~\ref{tab:cluster-balance}). Across all ten configurations, \sys{} reduces the cross-engine standard deviation of KV-cache occupancy by 44--79\% (e.g., $-68.0\%$ on Multitask, $-62.4\%$ on CodeAgent, $-72.5\%$ on the stretched tail, and $-68.6\%$ at 32B) and lowers single-engine peak KV-cache utilization by 13--35\%, directly mitigating memory-pressure hotspots on overloaded engines. On the 14B 4-engine workloads, the longest idle wait experienced by any engine (\emph{worst engine stall}) drops by 41--80\% (Multitask: $111.6 \to 66.2$\,s; CodeAgent: $300.2 \to 150.2$\,s; Straggler: $22.2 \to 5.8$\,s), and the standard deviation of engine finish times falls by 40--80\%. Relieving overloaded engines also accelerates in-flight requests before the final straggler finishes: on Multitask, p95 time-to-first-token (TTFT) drops by 15.2\% ($1.07 \to 0.91$\,s), p99 request completion latency drops by 4.1\% ($206.8 \to 198.3$\,s), and generation throughput rises by 4.1\% ($2{,}271 \to 2{,}364$\,tokens/s); on CodeAgent, p95 vLLM chunk latency drops by 10.2\% ($22.7 \to 20.4$\,s), p95 request latency drops by 5.8\% ($480.5 \to 452.6$\,s), and throughput rises by 4.2\%.
\begin{figure}[t]
  \centering
  \includegraphics[width=0.85\columnwidth]{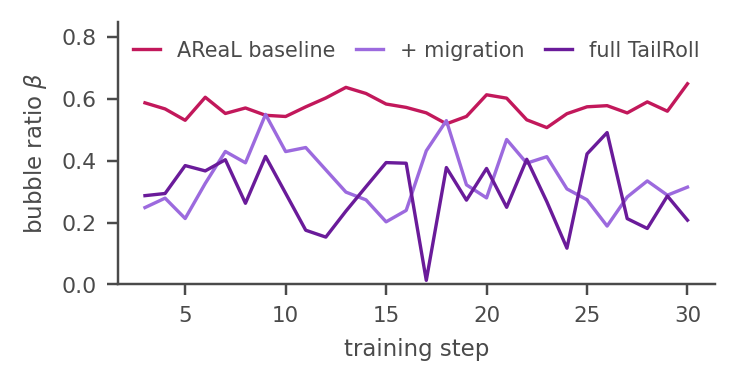}
  \caption{\scriptsize Per-step bubble ratio, Qwen3-32B multimodal-heavy ($N{=}0$). Mean $\beta$: $0.573 \to 0.340$ (migration) $\to 0.295$ (full \sys{}), $-$49\%.}
  \label{fig:eval-trend32b}
\end{figure}

\textbf{Results at larger model and cluster scales.} At Qwen3-32B (TP=4, 32 NPUs) the baseline wastes 57\% of engine-time and full \sys{} cuts that to 29.5\%---a 49\% reduction that holds step after step (Fig.~\ref{fig:eval-trend32b}). As on the short 14B multimodal traces, this reduction appears mainly as capacity rather than span: rollout time is 139.2\,s for the baseline and 140.1\,s for full \sys{} (Table~\ref{tab:summary}); migration evaluated on its own, without termination, reaches 136.9\,s. Scaling out to $E{=}8$ engines raises the baseline bubble (0.372, as more engines magnify imbalance) and full \sys{} removes 21\% of it while improving throughput 927 $\to$ 961 tokens/s ($+$3.7\%); the makespan-bound behavior of Sec.~\ref{sec:eval-bound} is therefore not an artifact of the 4-engine testbed.

\begin{figure}[t]
  \centering
  \includegraphics[width=0.85\columnwidth]{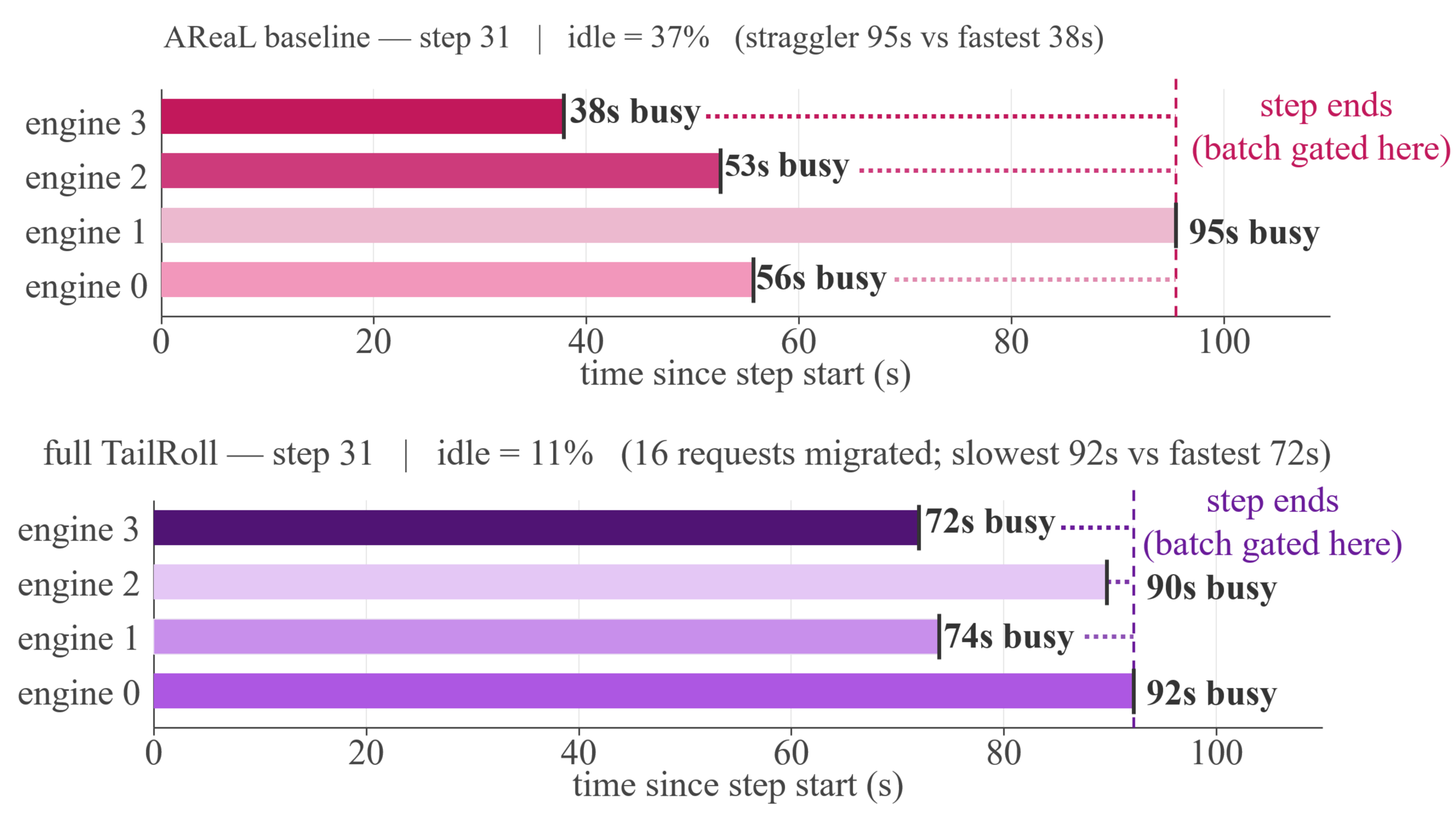}
  \caption{\scriptsize Per-engine busy span, one multimodal-heavy step ($N{=}0$, Qwen3-14B), step 31: \areal{} (top) vs.\ full \sys{} (bottom). 16 requests migrated off engine 1; idle time 37\% $\to$ 11\%, step $-$3.3\,s ($-$3.4\%).}
  \label{fig:onestep-compare}
\end{figure}

\textbf{Per-engine behavior in one step.}
The multimodal-heavy workload sits close to its makespan lower bound on most steps, leaving little room for span improvement (Table~\ref{tab:bound}, Sec.~\ref{sec:eval-bound}). To illustrate the mechanism we show one step from the same run with an unusually wide baseline imbalance---not a typical step. Fig.~\ref{fig:onestep-compare} shows it in action: in the baseline, three engines drain early and wait for the straggler, while \sys{} detects the widening freeness gap and rebalances so all four engines finish close together. Because this step has slack beyond the bound, span moves visibly here even though the row mean for this workload is flat ($+0.2\%$, Table~\ref{tab:levers}).
\subsection{Rollout-Time Gains on Real RL Tasks}
\label{sec:eval-realtask}
\label{sec:eval-amc12}
We now evaluate \sys{} on two workloads that train with a live reward signal and natural stopping, complementing the forced-length replays of Sec.~\ref{sec:eval-headline}. \emph{AMC12} is a fully native long-CoT RL workload with the model's own reasoning lengths and exact-match reward. \emph{Multimodal (live reward)} uses the multimodal trace for request structure with natural stopping (EOS honored, no forced output lengths), training with a live reward signal. For this multimodal live-reward workload, we report a 50-step matched-arm comparison and a longer 300-step training run (Sec.~\ref{sec:eval-setup}).
\begin{figure*}[t]
  \centering
  \begin{minipage}[t]{0.48\textwidth}
    \centering
    \includegraphics[width=0.95\linewidth]{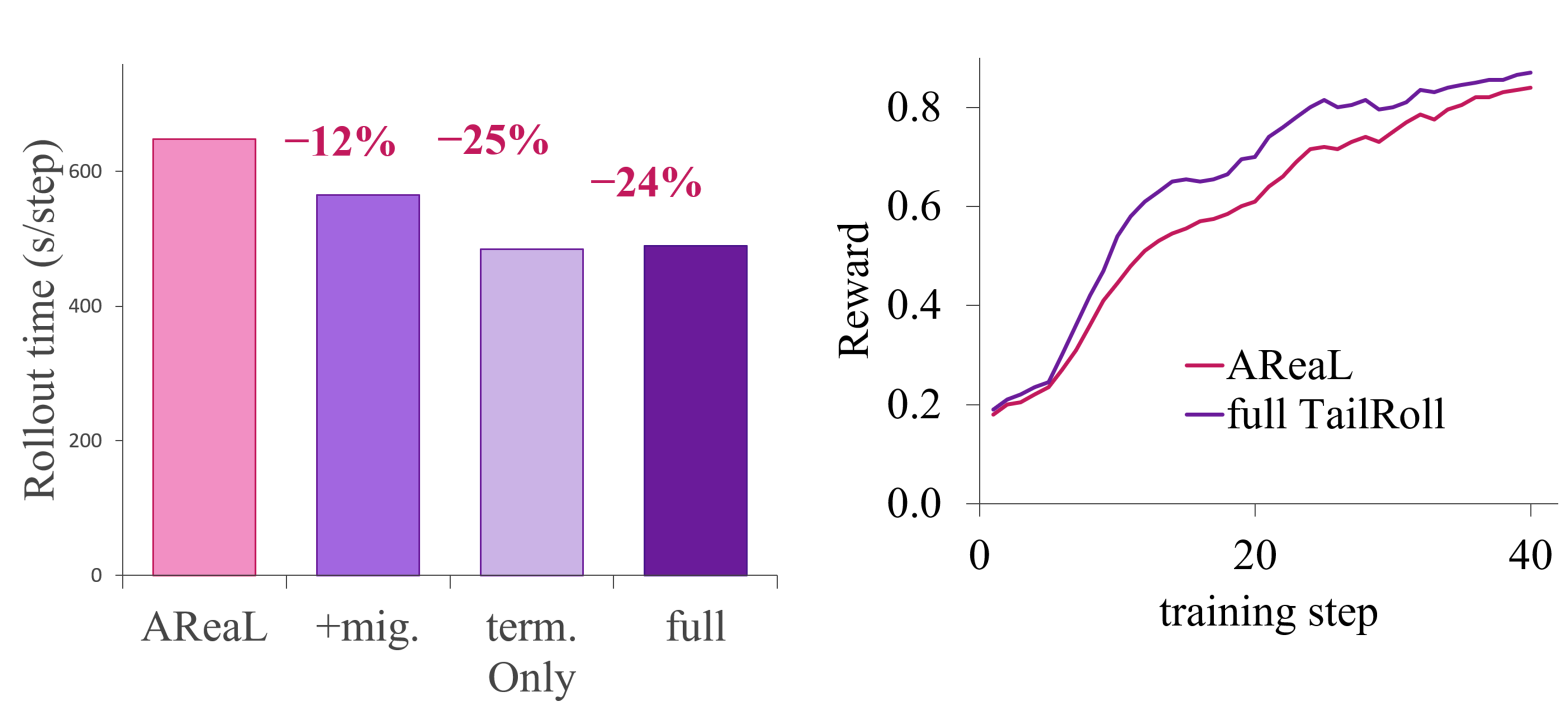}
    \caption{\scriptsize AMC12 long-CoT RL (Qwen3-14B, 8K, $n{=}8$, exact-match). \textbf{Left:} rollout time/step, four arms. \textbf{Right:} reward.}
    \label{fig:eval-amc12}
  \end{minipage}\hfill
  \begin{minipage}[t]{0.48\textwidth}
    \centering
    \includegraphics[width=0.95\linewidth]{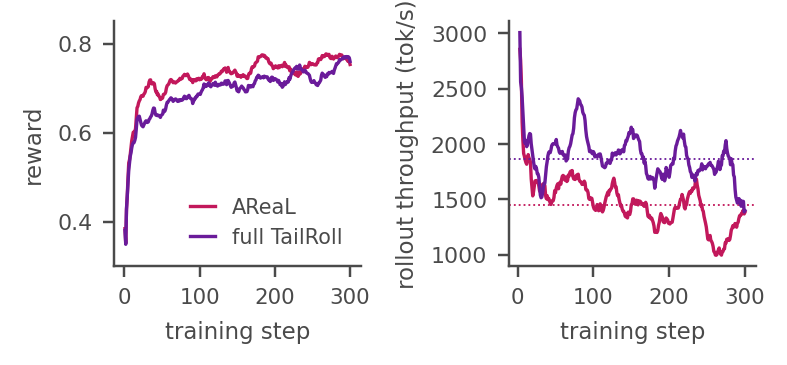}
    \caption{\scriptsize 300-step multimodal live-reward run. \textbf{Left:} reward. \textbf{Right:} rollout throughput ($+$29\% for \sys{}).}
    \label{fig:eval-realreward}
  \end{minipage}
\end{figure*}
\textbf{AMC12: shorter rollouts without reducing reward.} AMC12 runs Qwen3-14B with thinking enabled on real competition problems; the length tail is the model's own reasoning-length variance, and the workload sustains 270 requests concurrent during the longest job. At the 8K token budget, we compare four arms: the \areal{} baseline, migration alone, sufficiency-gated termination alone (5\% per-step cap), and full \sys{} (migration, tail-aware balancing, and termination together). Migration alone shortens rollout from 644.7 to 565.1\,s per step ($-$12.3\%) and cuts $\beta$ from 0.175 to 0.071 ($-$59\%), consistent with the deeply-batched regime of Sec.~\ref{sec:eval-bound}: moving load off the engine serving the critical-path request relieves its decode contention. Termination alone reaches a larger span reduction, $-$25.0\% (483.5\,s), by cutting stragglers directly. Full \sys{} combines both mechanisms and reaches $-$23.6\% rollout time (492.8\,s) and $-$21.1\% end-to-end training step time ($975.6 \to 769.6$\,s). At the request level, full \sys{} reduces median and p95 request completion latency by 33.4\% ($329.2 \to 219.1$\,s) and 28.0\% ($626.5 \to 451.3$\,s), cuts p95 vLLM chunk latency by 47.0\% ($64.6 \to 34.3$\,s), lowers p95 time-per-output-token (TPOT) by 27.8\% ($102 \to 74$\,ms) and p95 time-to-first-token (TTFT) by 88.4\% ($7.32 \to 0.85$\,s), and increases generation throughput by up to 12.8\% ($4{,}423 \to 4{,}991$\,tokens/s). That full \sys{} lands slightly below termination alone in span rather than above it reflects that the two mechanisms act on different parts of the distribution (Sec.~\ref{sec:eval-term} examines termination's operating range on its own); on a natural-stopping run the policy's own learning also shifts the length distribution between arms, so we do not expect the two levers to combine exactly additively here.

\textbf{AMC12 at a 16K token budget.} Raising the generation cap to 16K tokens exposes the model's unclipped reasoning tail ($p90{=}15{,}840$, $p99{=}16{,}384$ tokens) and raises rollout's share of the training step. Under this deeper tail, full \sys{} reduces rollout time by 33.9\% ($1{,}946.3 \to 1{,}287.1$\,s, saving 11 minutes per step) and cuts end-to-end RL training step time by 22.0\% ($3{,}134.9 \to 2{,}446.6$\,s). Simultaneously, $\beta$ falls by 73.7\% ($0.208 \to 0.055$), cross-engine KV-cache std drops by 78.8\% ($0.131 \to 0.028$), worst-engine stall time drops by 79.7\% ($803.2 \to 163.4$\,s), p95 request completion latency falls by 36.8\% ($1{,}850.7 \to 1{,}169.2$\,s), p95 TTFT falls by 70.7\% ($33.5 \to 9.8$\,s), p95 TPOT falls by 24.2\% ($143 \to 108$\,ms), and generation throughput increases by 33.3\% ($2{,}365 \to 3{,}153$\,tokens/s), while maintaining a higher mean task reward ($0.768$ vs.\ $0.730$, and $0.864$ vs.\ $0.841$ over the final 20 steps). In both the 8K and 16K runs, the accelerated arms reach comparable or higher measured reward: at 8K, final-40-step reward is 0.834 (full \sys{}) and 0.823 (termination alone), against 0.779 for the baseline (Fig.~\ref{fig:eval-amc12}, right). Separately from this observation, the sufficiency gate is designed so that a trajectory is only terminated when its group already holds $k_{\min}$ completed siblings, which by construction leaves every group with enough finished samples for an advantage estimate; the reward curves above show that, on this run, this design does not come at a measurable cost to training.

\textbf{Multimodal (live reward): a 50-step matched comparison.} On a 50-step pair with natural stopping (EOS honored) driven by the multimodal trace with a live training reward, \sys{} reduces rollout time per step from 15.9 to 10.5\,s ($-$34\%), $\beta$ from 0.191 to 0.088 ($-$54\%), worst-engine stall time by 66.0\% ($5.1 \to 1.7$\,s), and p99 request latency by 5.8\% ($7.53 \to 7.09$\,s), while increasing generation throughput by 18.4\% ($1{,}859 \to 2{,}200$\,tokens/s) at nearly identical mean reward (0.670 vs.\ 0.671) and similar tokens generated per step (20.3K vs.\ 20.1K).
\textbf{Multimodal (live reward): a 300-step training run.} To check whether the 50-step result holds over a longer run, we train the same matched pair for 300 steps (Fig.~\ref{fig:eval-realreward}). The two final-40-step reward means are nearly identical (0.761 baseline vs.\ 0.763 full \sys{}), satisfying the admissibility constraint of Eq.~\eqref{eq:objective}, while \sys{} sustains 29\% higher generation throughput and a 37\% lower bubble ratio (0.199 $\to$ 0.126) across the entire run. For an operator, the practical result is that \sys{} maintains comparable final reward while sustaining higher generation throughput and reducing rollout waste over the full training run.
\subsection{What Limits Rollout-Step Time}
\label{sec:eval-bound}
The waste decomposition of Sec.~\ref{sec:formulation} identifies the source of rollout waste and the scheduling lever that can address it. We validate it by computing, for each rollout step, a makespan lower bound $\mathrm{span}_{\min} = \max(\bar b, L)$ (Eq.~\eqref{eq:terms}), where $\bar b$ is the average work per engine and $L$ the longest request service time, both taken from the controller's recorded span records. This is a lower bound \emph{under the recorded request service times}: it is not a counterfactual system optimum, because migration can itself change $L$ by altering batch composition and thus per-request decode speed (as the second paragraph below shows).
\begin{table}[t]
\centering
\caption{\scriptsize Distance from makespan bound (Qwen3-14B, $N{=}0$); $\mathrm{span}_{\min}=\max(\bar b,L)$ (Eq.~\eqref{eq:terms}). Span is the engine window (first start to last finish), excluding trainer submission/collection (0.8--1.0\,s/step) included in Tables~\ref{tab:summary}--\ref{tab:levers}. Means over matched steps; ratio computed per step.}
\label{tab:bound}
\footnotesize
\setlength{\tabcolsep}{2.5pt}
\resizebox{0.92\columnwidth}{!}{%
\begin{tabular}{@{}llrrrr@{}}
\toprule
Workload & Arm & span & $\bar b$ & $L$ & \shortstack{span/\\$\mathrm{span}_{\min}$} \\
\midrule
\multirow{4}{*}{Multimodal-heavy} & baseline        & 89.2 & 74.0 & 88.5 & 1.008 \\
 & migration       & 88.7 & 80.5 & 87.8 & 1.010 \\
 & + balancing     & 89.4 & 83.5 & 88.5 & 1.009 \\
 & oracle dispatch & 89.1 & 73.6 & 88.3 & 1.009 \\
\midrule
Multimodal-normal & baseline & 57.5 & 37.0 & 56.5 & 1.020 \\
Straggler pool    & baseline & 28.7 & 13.4 & 28.1 & 1.045 \\
CodeAgent (ep.)   & baseline & 540.1 & 389.5 & 540.0 & 1.000 \\
Multimodal ($p99/p50{=}80$) & baseline & 201.4 & 153.4 & 200.7 & 1.003 \\
Multitask         & baseline & 236.6 & 175.5 & 236.6 & 1.000 \\
AMC12 (8K)        & baseline & 643.9 & 521.0 & 643.8 & 1.000 \\
AMC12 (16K)       & baseline & 1945.5 & 1530.1 & 1945.4 & 1.000 \\
\bottomrule
\end{tabular}}
\end{table}
\textbf{Rollout time is close to the makespan bound.} Across all reported configurations $\mathrm{span}/\mathrm{span}_{\min}$ ranges from 1.000 to 1.045 and is usually within 1\%: for the recorded request service times, the achieved rollout span is close to the scheduling lower bound, so placement alone leaves little room on the table. The mean $L$ exceeds the mean $\bar b$ in every row, indicating that the longest request, rather than the average engine load, sets the dominant term. Consistent with this, the waste decomposition attributes most of the measured waste to straggler waste $W_{\mathrm{strag}}$ rather than imbalance waste $W_{\mathrm{imb}}$ (Eq.~\eqref{eq:split}): on the heavy trace $W_{\mathrm{imb}}$ accounts for under 1\% of engine-time against 16\% for $W_{\mathrm{strag}}$, and on the straggler pool the fractions are 2\% and 51\%.

\textbf{The effective scheduling lever depends on batch depth.} Because the dominant term is $L$, the lever that shortens a step is the one that can shorten the longest request, and which lever that is depends on how deeply the tail is still batched. When the tail remains deeply batched, migration can shorten $L$ itself: moving requests off the binding engine relieves decode contention and speeds up the requests that remain---on our hardware the per-request decode rate roughly doubles as batch depth falls from 256 to 1. This explains the rollout-time reductions on AMC12 (concurrency 270) and on the CodeAgent episodes. Once the tail has drained to a single request, no co-batched requests remain whose migration could relieve its contention; the step stays bound by that request, and only termination can shorten the critical path, subject to the group-sufficiency condition (Sec.~\ref{sec:eval-term}).

\subsection{Dispatch-Time Placement vs.\ In-Flight Migration}
\label{sec:eval-baselines}
A natural objection is that a smarter \emph{router} could obtain these gains without touching running requests. We test this with an oracle dispatcher: at each dispatch it reads the incoming request's true output length and the true residual work of every in-flight request and uses them to pick the destination engine, but it cannot move a request once started. It is an optimistic reference for dispatch-time placement---supplied with more accurate workload information than any practical length predictor could provide (Sec.~\ref{sec:eval-setup})---rather than an achievable scheme.
\begin{table}[t]
\centering
\caption{\scriptsize Dispatch vs.\ in-flight migration (Qwen3-14B, multimodal-heavy, $N{=}0$).}
\label{tab:dispatch}
\scriptsize
\setlength{\tabcolsep}{2.5pt}
\resizebox{0.92\columnwidth}{!}{%
\begin{tabular}{@{}lrrr@{}}
\toprule
Policy & Rollout s/step & $\beta$ & Migr./step \\
\midrule
Round-robin (\areal{} stock)  & 90.0 & 0.171 & 0 \\
Least-loaded dispatch         & 91.1 & 0.175 & 0 \\
Oracle-work dispatch          & 89.8 & 0.173 & 0 \\
\midrule
Round-robin + migration (\sys{}) & 89.4 & \textbf{0.092} & 14.8 \\
Oracle-work + migration       & \textbf{87.4} & 0.094 & 14.2 \\
\bottomrule
\end{tabular}}
\end{table}
\textbf{Dispatch-time placement provides limited improvement.} The dispatch-only policies produce similar results (Table~\ref{tab:dispatch}): round-robin, least-loaded, and the oracle all span 89.8--91.1\,s at $\beta$ 0.171--0.175. Even with true request-length information, oracle placement does not substantially reduce either metric, because the waste is created \emph{after} dispatch, as decode durations diverge---the direct confirmation of our motivation study (Sec.~\ref{sec:motivation}) and the clearest argument for acting during decode rather than at the router.

\textbf{In-flight migration reduces waste under different dispatch policies.} Combining migration with round-robin and oracle-work dispatch drives $\beta$ to 0.092--0.094, a 46\% reduction relative to the corresponding dispatch-only configuration, and oracle placement combined with migration also yields the shortest rollout time (87.4\,s). In-flight migration thus addresses runtime imbalance that dispatch-time placement cannot resolve once requests are running, and it composes with more informed dispatch policies rather than replacing them. The same pattern holds at 32B, where oracle dispatch alone barely changes the bubble ratio (0.572 against 0.573 for the stock dispatcher), whereas migration combined with the stock dispatcher reduces it to 0.340 (Fig.~\ref{fig:eval-trend32b}; no oracle-plus-migration run was performed at 32B).
\subsection{Policy Ablations: Detection Signal, Cost Gate, and Tail Balancing}
\label{sec:eval-policy}
\begin{figure}[t]
  \centering
  \includegraphics[width=0.85\columnwidth]{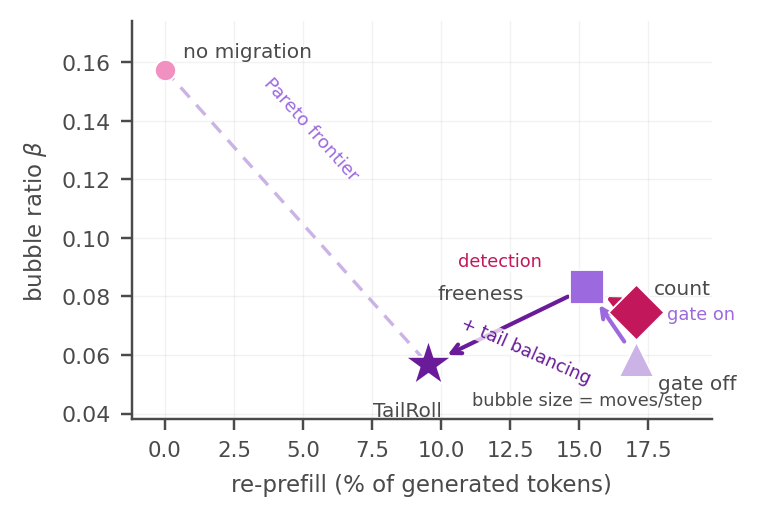}
  \caption{\scriptsize Policy-cost ablation, multimodal-heavy ($N{=}0$, Qwen3-14B; steps 3--30; termination disabled). $\beta$ vs.\ re-prefill cost; marker area = moves/step (lower-left is better).}
  \label{fig:eval-policycost}
\end{figure}

We ablate the three policy components on multimodal-heavy: the detection signal (the freeness score of Sec.~\ref{sec:detect} or Seer's running-request count spread), the cost-benefit gate (Inequality~\eqref{eq:cb}), and tail-aware balancing (Sec.~\ref{sec:spread}). Fig.~\ref{fig:eval-policycost} shows the trade-off between bubble ratio and re-prefill cost across policy variants, with bubble area representing the number of moves. The arrows indicate three comparisons: count-based versus freeness-based detection, the cost-benefit gate enabled versus disabled, and tail-aware balancing disabled versus enabled. Table~\ref{tab:policy-ablation} reports the exact bubble ratio, rollout time, migration count, and re-prefill cost. Since these ablation runs cover 30 steps while the main experiments cover 50, all values are computed over the common horizon (steps 3--30).
\begin{table}[t]
\centering
\caption{\scriptsize Policy ablation, multimodal-heavy ($N{=}0$, Qwen3-14B; steps 3--30). Moves/step = general $+$ tail moves; \sys{} row: 11.2 general $+$ 1.8 tail.}
\label{tab:policy-ablation}
\footnotesize
\setlength{\tabcolsep}{2.5pt}
\resizebox{0.92\columnwidth}{!}{%
\begin{tabular}{@{}l|c|c|c|r|r|r|r@{}}
\toprule
Variant & Detection & Gate & \shortstack{Tail\\bal.} & \shortstack{Rollout\\s/step} & $\beta$ & \shortstack{Moves/\\step} & \shortstack{Re-prefill\\\%} \\
\midrule
No migration       & ---      & --- & --- & 91.1 & 0.157 & 0    & 0 \\
Count signal       & count    & on  & off & 89.2 & 0.075 & 21.8 & 17.0 \\
Freeness           & freeness & on  & off & 90.2 & 0.083 & 15.1 & 15.3 \\
Freeness, gate off & freeness & off & off & 90.5 & 0.059 & 15.3 & 17.1 \\
\sys{} policy      & freeness & on  & on  & 91.2 & \textbf{0.057} & 13.0 & \textbf{9.5} \\
\bottomrule
\end{tabular}}
\end{table}
\textbf{Effect of the detection signal.}
With the cost gate enabled and tail balancing disabled, count-based and freeness-based detection achieve comparable bubble ratios (0.075 and 0.083, respectively). However, freeness-based detection requires 31\% fewer moves (15.1 vs.\ 21.8 per step) and 10\% less re-prefill (15.3\% vs.\ 17.0\%). Thus, the main benefit of the freeness signal is not a lower bubble ratio by itself, but achieving a similar reduction with fewer migrations and less recomputation. It identifies engines with genuinely available capacity rather than relying only on differences in running-request counts.

\textbf{Effect of the cost-benefit gate.}
With freeness detection and tail balancing disabled, turning off the gate lowers $\beta$ from 0.083 to 0.059, but increases re-prefill from 15.3\% to 17.1\%. The number of moves remains nearly unchanged (15.1 vs.\ 15.3 per step). Thus, disabling the gate permits moves that reduce the bubble ratio but incur more recomputation. The gate rejects candidates whose expected reduction in remaining decode time does not justify their re-prefill cost, trading some bubble reduction for lower migration overhead.

\textbf{Effect of tail-aware balancing.}
Adding tail-aware balancing to the freeness-based policy with the cost gate enabled reduces $\beta$ from 0.083 to 0.057 and re-prefill from 15.3\% to 9.5\%. The total number of moves also decreases from 15.1 to 13.0 per step. In the full policy, these comprise 11.2 general migrations and 1.8 tail-balancing moves. Thus, tail-aware balancing reduces the need for general migrations while continuing to address the drained tail. The full policy achieves the lowest re-prefill cost in the table and a bubble ratio comparable to the ungated variant (0.057 vs.\ 0.059), with 44\% less recomputation. Compared with count-based detection, it reduces re-prefill by 44\% and total moves by 40\%, while also achieving a lower bubble ratio (0.057 vs.\ 0.075). Rollout time remains within 89.2--91.2\,s across the migrating variants, consistent with Sec.~\ref{sec:eval-bound}: on this workload, the longest request bounds the step, so differences between policies appear mainly in bubble ratio and recomputation cost rather than rollout span.
\subsection{Sensitivity to Staleness, Group Size, and Batch Size}
\label{sec:eval-sensitivity}
\begin{figure}[t]
  \centering
  \includegraphics[width=0.85\columnwidth]{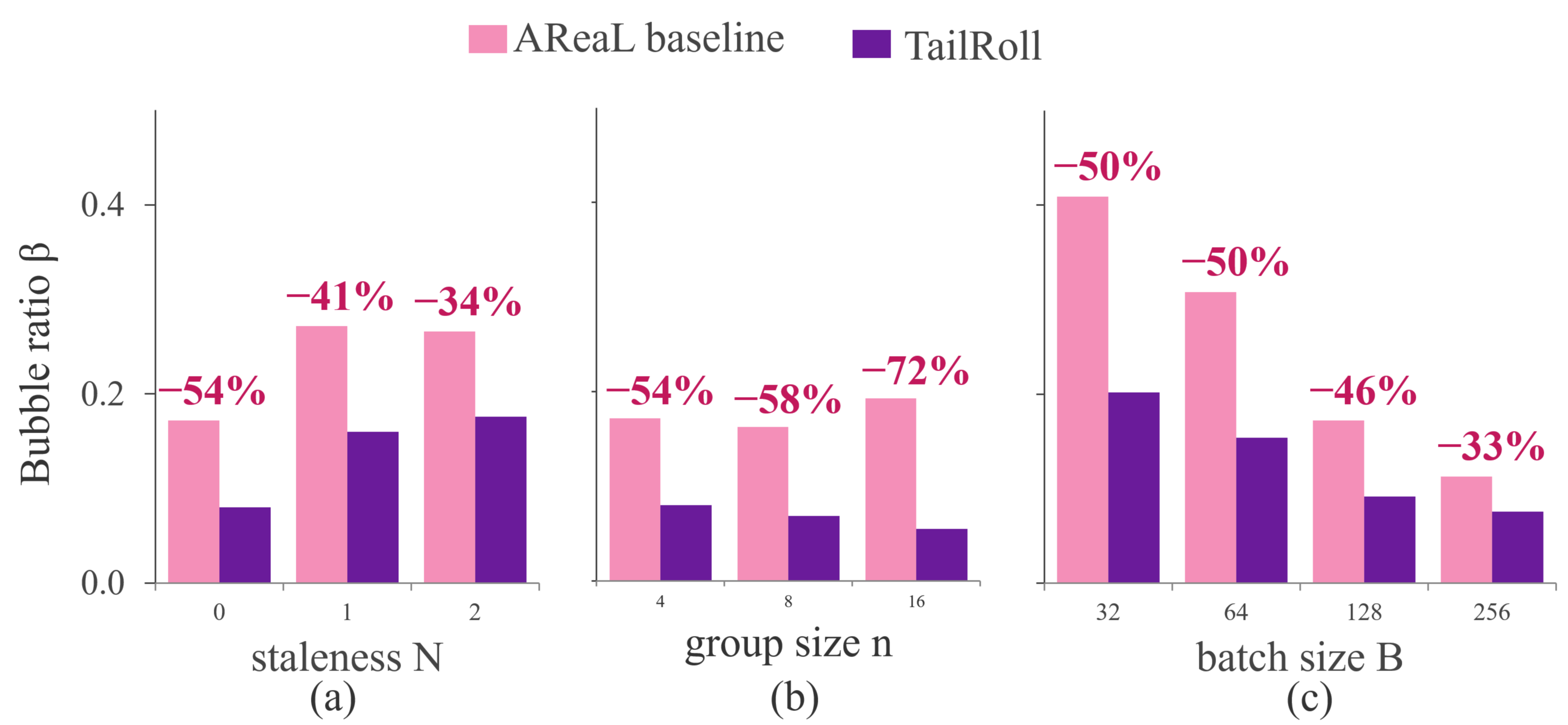}
  \caption{\scriptsize Bubble ratio vs.\ (a) staleness $N$, (b) group size $n$, (c) batch size $B$. (a), (b) full \sys{}; (c) migration only. Percentages: reduction vs.\ baseline.}
  \label{fig:eval-sensitivity}
\end{figure}

We examine how the effectiveness of \sys{} changes with three parameters: training staleness $N$, group size $n$, and batch size $B$. Fig.~\ref{fig:eval-sensitivity} compares the bubble ratio of the \areal{} baseline and \sys{} as each parameter varies. The percentages above the bars report the relative reduction in bubble ratio achieved by \sys{}.
\textbf{The reduction persists across staleness settings.}
As $N$ increases from 0 to 2, \sys{} continues to reduce the bubble ratio, with reductions of 54\%, 41\%, and 34\%, respectively (Fig.~\ref{fig:eval-sensitivity}(a)). Although the relative benefit decreases with increasing staleness, the reduction remains substantial across all three settings.

\textbf{The benefit holds across group sizes.}
For group sizes of 4, 8, and 16, \sys{} reduces the bubble ratio by 54\%, 58\%, and 72\%, respectively (Fig.~\ref{fig:eval-sensitivity}(b)). The largest reduction occurs at $n{=}16$, showing that the scheduling improvements persist as the group size increases.

\textbf{The improvement persists across batch sizes.}
As batch size increases from 32 to 256, the bubble-ratio reduction changes from 50\% to 33\%, with reductions of 50\% and 46\% at batch sizes 64 and 128, respectively (Fig.~\ref{fig:eval-sensitivity}(c)). Thus, \sys{} continues to reduce rollout waste across the tested batch sizes, although the relative reduction is smaller at larger batch sizes.
\subsection{Termination: Reward Impact and Straggler Reduction}
\label{sec:eval-term}
Termination directly removes remaining work from straggler trajectories, including when the tail has drained and migration can no longer relieve decode contention. We evaluate it along two axes: its effect on reward and its effect on rollout time.

\textbf{Sufficiency-gated termination maintains reward quality in the tested tasks.} A trajectory may be terminated only after $k_{\min}{=}3$ trajectories of its group have completed; with the default $n{=}4$, the one remaining trajectory can then be terminated, and its partial output is masked from the loss. On the 300-step multimodal live-reward run (Fig.~\ref{fig:eval-realreward}), baseline and full \sys{} reach nearly identical final-40-step reward (0.761 vs.\ 0.763), but this pair does not isolate termination from migration and balancing. On AMC12 ($n{=}8$, 40 steps, one seed), the termination-only configurations with 2\% and 5\% caps reach higher final-40-step rewards than the baseline (0.830 and 0.823 vs.\ 0.779) while reducing rollout time by 18.9--25.0\%. These results show that the sufficiency condition allows \sys{} to shorten straggler trajectories without reducing the measured reward in either task; they do not by themselves establish that termination leaves learning unaffected in general.

\textbf{The termination cap controls the trade-off between rollout time and reward.} On AMC12, a 2\% per-step cap reduces rollout time by 18.9\% relative to the baseline, and the default 5\% cap by 25.0\%. Increasing the cap to 10\%, however, lengthens rollout time to 9.3\% \emph{above} the baseline (703.7 vs.\ 643.9\,s) and lowers reward to 0.758. We did not isolate the cause of this regression; the lower reward is consistent with overly aggressive termination removing long trajectories that still carry useful training signal. The 5\% cap gives the better balance between rollout acceleration and reward in this experiment. The sufficiency gate ensures that each group retains enough completed samples, whereas the cap bounds how aggressively trajectories are terminated. On the short-step replayed traces, the termination gate fires infrequently (0.02--0.15 times per step) and reduces the total number of generated tokens by at most 0.015\% (Sec.~\ref{sec:eval-overhead}). These workloads rarely contain a straggler whose group already has enough completed samples, so termination remains inactive when its expected benefit is limited.

\subsection{Overhead}
\label{sec:eval-overhead}
\sys{} introduces four costs: (i) chunked HTTP round trips, paid by every request; (ii) controller polling; (iii) re-prefill of migrated contexts; and (iv) tokens generated by trajectories that are later terminated. Table~\ref{tab:overhead} groups them into runtime overhead, additional execution work, and mechanism activity; the chunking cost comes from a single-engine microbenchmark and the rest from the run logs. The one-time startup probe (Sec.~\ref{sec:impl-health}) takes 2.3\,s per run and is not included.
\begin{table}[t]
\centering
\caption{\scriptsize Mechanism overhead (Qwen3-14B, multimodal-normal/heavy, $N{=}0$, full \sys{}). Load-monitor CPU: polling thread incl.\ policy; incremental = vs.\ baseline (monitor on, mechanisms off).}
\label{tab:overhead}
\small
\setlength{\tabcolsep}{3pt}
\resizebox{0.92\columnwidth}{!}{%
\begin{tabular}{@{}llrr@{}}
\toprule
Category & Metric & Normal & Heavy \\
\midrule
\multirow{4}{*}{Runtime overhead}
 & Chunking, added wall-clock vs.\ unchunked (\%) & 0.2 & 0.2 \\
 & Load-monitor CPU time (s/step)                & 0.51 & 0.63 \\
 & \quad of which incremental over baseline (s/step) & 0.12 & 0.23 \\
 & Load-monitor CPU time / rollout time (\%)     & 0.9 & 0.7 \\
\midrule
\multirow{3}{*}{Additional execution work}
 & Re-prefill tokens / step                      & 4\,844 & 16\,470 \\
 & Re-prefill (\% of generated tokens)           & 5.2 & 9.7 \\
 & Loss-masked partial tokens / step             & 167 & 49 \\
\midrule
\multirow{3}{*}{Mechanism activity}
 & Migrations / step                             & 4.1 & 11.5 \\
 & Spread requests / step                        & 3.4 & 1.6 \\
 & Terminations / step                           & 0.06 & 0.02 \\
\bottomrule
\end{tabular}}
\end{table}
\textbf{Runtime overhead.} Chunking's 0.2\% wall-clock cost comes from a single-engine microbenchmark: with $C{=}512$, a 3072-token budget crosses five chunk boundaries, each one additional HTTP round trip, with the prefix cache avoiding recomputation (Sec.~\ref{sec:mechanism}). Controller polling (Table~\ref{tab:overhead}) is a CPU resource cost, not added wall-clock latency; the baseline's own monitor uses 0.39--0.40\,s of that total. With 8 engines the total is 0.35\,s per step (0.4\%) and with Qwen3-32B it is 0.31\,s (0.2\%), so polling cost does not increase with engine count in the tested configurations.

\textbf{Additional execution work.} Re-prefill (Table~\ref{tab:overhead}) is the largest term, but because prefill runs roughly 
an order of magnitude faster per token than decode ($\sigma \approx 10$, Eq.~\eqref{eq:cost}), it translates to 
only a few additional engine-seconds per step; the cost-benefit gate reduces it by about 44\% relative to the ungated
 policy (Sec.~\ref{sec:eval-policy}). Terminated trajectories' loss-masked partial tokens are generated but excluded from
  the training loss---distinct from the net change in total generated-token count across configurations, which remains 
  within 0.015\% on the replayed traces (below).

\textbf{Migration preserves generated-token counts; termination shortens trajectories by design.} On the replayed traces, the migration and tail-aware-balancing arms generate exactly the same total number of output tokens as the baseline. Terminated trajectories are intentionally shorter: with termination enabled, total output tokens fall by at most 0.015\% (29\,245.3 vs.\ 29\,249.5 tokens per step on the straggler pool), and their partial outputs are masked from the training loss.

%% file: sections/8.Related-work.tex
\section{Related Work}
    \label{sec:related}

\begin{table}[!t]
    \centering
    \caption{\scriptsize Comparison of related systems with \sys{}.}
    \label{tab:related}
    \scriptsize
    \setlength{\tabcolsep}{2.5pt}
    \renewcommand{\arraystretch}{1.0}
    
    \resizebox{0.92\columnwidth}{!}{%
    \begin{tabular}{@{}lllcccccc@{}}
    \toprule
    \textbf{System} & \shortstack{\textbf{Sched.}\\\textbf{signal}} & \shortstack{\textbf{Place-}\\\textbf{ment}} &
    \shortstack{\textbf{Mid-dec.}\\\textbf{migr.}} &
    \shortstack{\textbf{Cost-aw.}\\\textbf{migr.}} &
    \shortstack{\textbf{Imbal.}\\\textbf{relief}} &
    \shortstack{\textbf{Tail}\\\textbf{relief}} &
    \shortstack{\textbf{Group-suff.}\\\textbf{stop}} &
    \shortstack{\textbf{Async}\\\textbf{RL}} \\
    \midrule
    \multicolumn{9}{@{}l}{\emph{Dispatch-time scheduling}} \\
    SortedRL~\cite{zhang2026sortedrl}   & Pred.\ len. & Batch & \xmark & -- & \cmark & \xmark & \xmark & \xmark \\
    RollPacker~\cite{gao2025rollpacker} & Pred.\ len. & Batch & \xmark & -- & \cmark & \xmark & \xmark & \xmark \\
    RhymeRL~\cite{he2025rhymerl}        & Pred.\ len. & Batch & \xmark & -- & \cmark & \xmark & \xmark & \xmark \\
    \midrule
    \multicolumn{9}{@{}l}{\emph{Asynchronous / disaggregated RL}} \\
    \areal{}~\cite{fu2025areal}         & --    & --    & \xmark & -- & \xmark & \xmark & \xmark & \cmark \\
    AsyncFlow~\cite{han2025asyncflow}   & --    & --    & \xmark & -- & \xmark & \xmark & \xmark & \cmark \\
    StreamRL~\cite{zhong2025streamrl}   & --    & --    & \xmark & -- & \xmark & \xmark & \xmark & \cmark \\
    \midrule
    \multicolumn{9}{@{}l}{\emph{Call-boundary rebalancing (agentic)}} \\
    Heddle~\cite{zhang2026heddle}       & Obs.  & Call  & \xmark & -- & \cmark & \xmark & \xmark & \cmark \\
    Dressage~\cite{dressage2026stepbalance} & Obs. & Call & \xmark & -- & \cmark & \xmark & \xmark & \cmark \\
    \midrule
    \multicolumn{9}{@{}l}{\emph{Partial / early-stopped rollouts}} \\
    APRIL~\cite{zhou2025april}, Kimi k1.5~\cite{team2025kimi} & -- & -- & \xmark & -- & \xmark & \cmark & \xmark & \xmark \\
    verl partial rollout~\cite{sheng2025hybridflow} & -- & -- & \xmark & -- & \xmark & \cmark & \xmark & \cmark \\
    Selective Rollout / ESPO~\cite{Zheng2026ActOnlyWhenItPays,Li2026ESPO} & -- & -- & \xmark & -- & \xmark & \cmark & \xmark & \xmark \\
    \midrule
    \multicolumn{9}{@{}l}{\emph{In-flight migration}} \\
    Llumnix~\cite{sun2024llumnix}       & Obs. & KV blk & \cmark & \xmark & \cmark & \xmark & \xmark & -- \\
    Seer~\cite{qin2026seer}             & Obs.\ (count) & Chunk & \cmark & \xmark & \cmark & \xmark & \xmark & \xmark \\
    \textbf{\sys{}}                     & \textbf{Obs.\ (freeness)} & \textbf{Chunk} & \cmark & \textbf{\cmark} & \cmark & \cmark & \cmark & \cmark \\
    \bottomrule
    \end{tabular}%
    }
    
    \vspace{2pt}
    \begin{minipage}{\columnwidth}
    \raggedright\scriptsize
    Sched. signal $=$ scheduling signal; Placement $=$ placement unit; Mid-dec.\ migr.\ $=$ mid-decode migration; Cost-aw.\ migr.\ $=$ cost-aware migration; Imbal.\ $=$ imbalance; Group-suff.\ stop $=$ group-sufficiency stopping. \cmark/\xmark denote whether a system provides a capability; -- means not applicable.
    \end{minipage}
    \end{table}
    
\textbf{In-flight migration.}
Llumnix~\cite{sun2024llumnix} and Seer~\cite{qin2026seer} also migrate decoding requests across instances (Sec.~\ref{sec:bg-migration}). \sys{} uses chunk-level migration like Seer but targets a broader rollout optimization problem: its decisions use engine freeness (running-request counts with KV-cache utilization) and a cost-benefit gate for re-prefill overhead; it supports bounded-staleness asynchronous training rather than only Seer's synchronous setting; and it combines migration with tail-aware load balancing and sufficiency-gated termination, addressing both imbalance and straggler-tail waste within a single rollout controller.

\textbf{Dispatch-time scheduling.} SortedRL~\cite{zhang2026sortedrl} reorders prompts by predicted output length, RollPacker~\cite{gao2025rollpacker} isolates predicted-long prompts into dedicated rounds, and RhymeRL~\cite{he2025rhymerl} routes predicted outliers using cross-epoch history. All place work according to a \emph{prediction} made before generation; Insight~3 shows the waste they target is created \emph{after} dispatch, once requests are already running. The two lines compose: prediction lowers the expected imbalance, migration removes the residual that prediction misses.
    
\textbf{Asynchronous and disaggregated RL.} \areal{}~\cite{fu2025areal}, AsyncFlow~\cite{han2025asyncflow}, StreamRL~\cite{zhong2025streamrl}, and RolloutPipe~\cite{rolloutpipe2026} overlap rollout with training or stream trajectories to a disaggregated trainer. This hides the cross-iteration bubble but not the within-window one: Insight~4 shows idle time persists inside each collection window at every staleness we measured.
    
\textbf{Runtime rebalancing at call boundaries.} Heddle~\cite{zhang2026heddle} migrates agentic trajectories between tool calls; Dressage's Step Balance~\cite{dressage2026stepbalance} re-routes each generation step of an agentic session from engine-load snapshots. Both share our diagnosis---fixed affinity cannot track runtime load---and both hand off tokens rather than KV, as we do. Their scheduling unit is the boundary \emph{between} calls, so a dispatched call runs to completion where it landed; on single-call rollouts there are no such boundaries and the mechanism reduces to least-loaded dispatch, which Sec.~\ref{sec:eval-baselines} measures.
    
\textbf{Partial and early-stopped rollouts.} APRIL~\cite{zhou2025april}, Kimi~k1.5~\cite{team2025kimi}, and verl's partial-rollout recipe~\cite{sheng2025hybridflow} interrupt unfinished trajectories and carry them into a later step, so the resumed segment is generated under newer weights; \sys{}'s migration finishes the trajectory in the same step under the same weights, and its handoff reuses the same pause-and-resume primitive these systems rely on. Selective Rollout and ESPO~\cite{Zheng2026ActOnlyWhenItPays, Li2026ESPO} stop or skip samples by a fixed heuristic on length, reward, or entropy. \sys{}'s termination differs in both trigger and gate: the trigger is a \emph{system} condition (concurrency has fallen into the step's late tail and a straggler is pinning it, Sec.~\ref{sec:termination}) and the gate is a \emph{live training} condition ($k_{\min}$ completed siblings in the group).
    
\textbf{Decoding acceleration.} TLT~\cite{hu2026taming} and RhymeRL-HistoSpec~\cite{he2025rhymerl} speed up the straggler with speculative decoding warmed on idle GPUs. They change how fast a request decodes; \sys{} changes where, and whether it must finish. The approaches are orthogonal and compose.

%% file: sections/9.Discussion-Conclusion.tex
\section{CONCLUSIONS AND FUTURE WORK}
\label{sec:discussion}
\label{sec:conclusion}

Rollout idle time in RL post-training has two sources: engine load imbalance ($W_{\mathrm{imb}}$) and straggler-induced critical-path delays ($W_{\mathrm{strag}}$). Separating them explains the limitations of dispatch-time placement and motivates workload-aware runtime scheduling. We proposed \sys{}, an in-flight scheduling framework integrated into \areal{} for synchronous and asynchronous RL training. It combines freeness-guided, cost-aware chunk migration and tail-aware balancing to reduce imbalance, with group-sufficiency termination to stop stragglers once enough GRPO siblings have completed. Across Qwen3-14B and Qwen3-32B workloads, \sys{} reduces rollout bubble ratio by 21--74\% and cross-engine KV-cache imbalance by 44--79\%, shortens rollout time by 24--34\% and end-to-end training step time by 21--22\%, and increases generation throughput by up to 33\%, while preserving training reward over long runs. Future work spans five directions: (1)~direct KV-cache transfer over high-bandwidth interconnects to avoid re-prefill for long-context and multi-turn agent trajectories, letting the scheduler choose between transfer and re-prefill by cost; (2)~extending \sys{} to vision-language models, balancing vision encoding, text prefill, and decoding across their distinct resource demands; (3)~using reward dispersion among completed GRPO siblings to improve termination decisions beyond completion counts; (4)~adaptive termination parameters ($k_{\min}$ and the per-step cap) driven by online trajectory-length percentiles and gate firing rates, which our 300-step runs did not directly measure; and (5)~multi-seed replication across the full experimental matrix to strengthen the evaluation beyond the within-run variability reported here.